\documentclass[twocolumn,trackchanges]{aastex7}

\def\Msun{\rm{\hbox{$\,$M$_{\odot}$}}}
\gdef\etal{{et al.\ }} 

\gdef\eg{{\it e.g.,\ }}
\gdef\gappr{\hbox{$_>\atop{^\sim}$}}
\revised{January 21, 2026}
\submitjournal{The Astrophysical Journal}
\begin{document}

\title{Stellar Mass Growth in the First Galaxies: Theory and Observation}

\author[orcid=0000-0002-6317-0037]{Alan Dressler}
\affiliation{Carnegie Science Observatories}
\email[show]{dressler@carnegiescience.edu}  

\author[orcid=0000-0000-0000-0002]{Andrew Benson} 
\affiliation{Carnegie Science Observatories}
\email{abenson@carnegiescience.edu}

%% Use the \collaboration command to identify collaborations. This command
%% takes an optional argument that is either a number or the word "all"
%% which tells the compiler how many of the authors above the command to
%% show. For example "\collaboration[all]{(DELVE Collaboration)}" wil include
%% all the authors above this command.
%%
%% Mark off the abstract in the ``abstract'' environment. 
\begin{abstract}
We compare the growth in stellar mass of galaxies in the $6<z<12$ epoch \citep{2024ApJ...964..150D} with predictions of a semi-analytic galaxy formation model---{\sc Galacticus} \citep{2012NewA...17..175B}. In contrast to diverse and controversial results that compare models and data for the \emph{luminosity} evolution of galaxies---reported in an abundance of recent papers, we find very good, unambiguous agreement in the more fundamental quantity of stellar mass---measured from \emph{JWST} observations--- and {\sc Galacticus} predictions. Specifically, we find good agreement for the \emph{shape} of the integrated stellar mass as a function of redshift without any adjustment of parameters, and in \emph{amplitude} as well, when `feedback' is lowered by a factor of 3 compared to that required to match later-universe models and data.  This result emerged from detailed investigation of the claim by Dressler \etal that bursts of star formation dominated the growth in stellar mass, specifically, that half of the galaxies with stellar mass \emph{growth} of  $M_* > 2\times10^8$\Msun\ in the epoch $8<z<6$ had less than $M_*<\times10^8$\Msun\ \emph{prior} to $z = 8$. Here too we find agreement between models and data, namely that these $\gappr 100$  Myr `bursts’  had strong \emph{in situ} growth at $z\le8$, or showed (in {\sc Galacticus}) substantial stellar and/or gas-rich mergers, and 30--40 Myr `starbursts’ as are common in $z<3$ galaxies.  We note that, if a theoretical simulation is unable to pass the test of matching the growth of \emph{stellar mass}, any success in reproducing the luminosity function is meaningless.
\end{abstract}

%% Keywords should appear after the \end{abstract} command. 
%% The AAS Journals now uses Unified Astronomy Thesaurus (UAT) concepts:
%% https://astrothesaurus.org
%% You will be asked to selected these concepts during the submission process
%% but this old "keyword" functionality is maintained in case authors want
%% to include these concepts in their preprints.
%%
%% You can use the \uat command to link your UAT concepts back its source.

%% From the front matter, we move on to the body of the paper.
%% Sections are demarcated by \section and \subsection, respectively.
%% Observe the use of the LaTeX \label
%% command after the \subsection to give a symbolic KEY to the
%% subsection for cross-referencing in a \ref command.
%% You can use LaTeX's \ref and \label commands to keep track of
%% cross-references to sections, equations, tables, and figures.
%% That way, if you change the order of any elements, LaTeX will
%% automatically renumber them.

\section{Introduction} 

The crucial step in cosmic history that led to our universe of great complexity was the birth of the first stars, whose lives and deaths produced chemical elements more complex than Lithium, the heaviest element made in the Big Bang. Even today, after 13 billion years of star formation, heavy elements account for only $\sim$2\%\ of the \emph{baryonic} mass (see Grevesse \& Sauva in \citealt{2000eaa..book.....M}), and in the first 100 million years, a hundred times less overall. Nevertheless, these heavier atoms, and the stars that made them, were the sole difference between a simple, sterile universe expanding into oblivion and the one that we inhabit---and inhabits us---of astonishing variety, complexity, and possibility. 

Our knowledge of physics has advanced to the point where we can aspire to examine and understand this essential step that made our existence and all that we know possible.

The James Webb Space Telescope was designed and built for this challenge \citep{1996STIN...9713870D}: to reach far enough, faint enough, and ``red enough'' to observe the first generation of stars---bound by the millions into \emph{the first galaxies}---to see how they formed and grew over the first billion years.  It is certain that even a mature study with \emph{JWST} of this epoch and its workings will fall short of revealing all, but for the first time, we are ``on site.''

Our theoretical models have evolved in tandem with this challenge, generating early-universe environments and simulating how dark and baryonic matter developed habitats for the first and second generations of stars, within the matrix of dark matter swells that would become the homes of galaxies \citep{2015ARA&A..53...51S}. In this study, our goal is to compare the growth of structure---both in dark and baryonic matter---and make the first comparison between theoretical models and what \emph{JWST's} cameras and spectrographs reveal.

This sounds like a task devoted to comparing the light coming from stars and galaxies, and comparing the earliest galaxies observed to those created in simulations.  It is perhaps unsurprising, then, that the voluminous literature---already many hundreds of papers (for a recent review, see \citealt{2025arXiv250117078S})---compares the \emph{light} we observe to what the models predict. However, this task is complicated by the fact that not just the births of stars, but also mergers, dust, and AGN, lead to so many possibilities that the process can come down to tweaking the models to see if they can reproduce what is observed, a disappointing substitute for a true model prediction that is then confirmed by data.

\begin{figure*}

\includegraphics[scale=0.85]
 {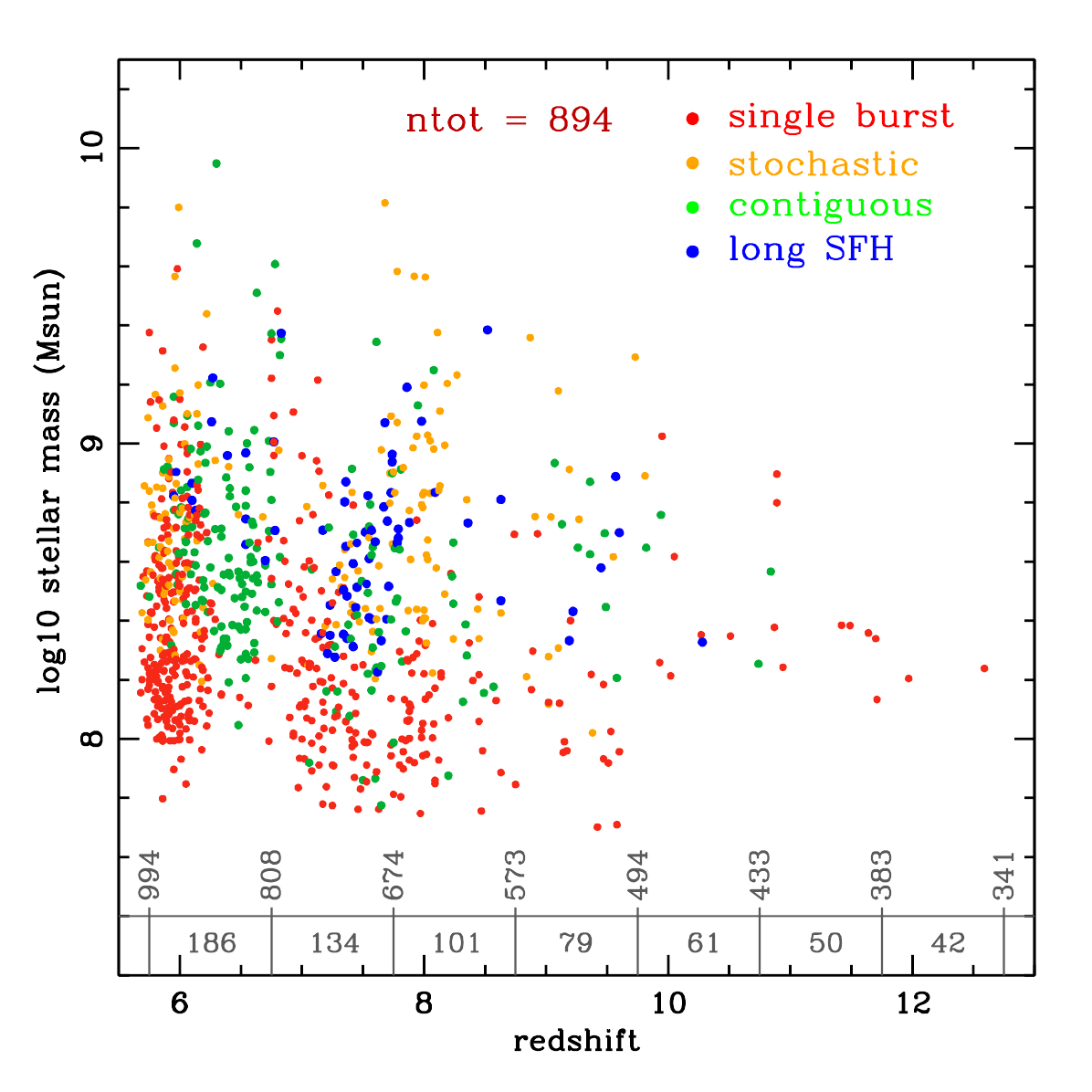}
\caption{The demographics of star formation histories represented as the growth of stellar mass in the first billion years of cosmic time. Circles show individual galaxies at their observed epochs from \citet{2024ApJ...964..150D}. The color of each point indicates its classification based on its star formation history as defined by \citet{2024ApJ...964..150D} and as indicated in the panel. Grey vertical numbers along the bottom axis show the age of the universe in Myr, while gray horizontal numbers show the duration in Myr between each identified epoch. The seven epochs shown cover $\sim$1~Gyr of cosmic time, in intervals that grow from $\sim$40~Myr at $z=12$ to almost $\sim$200~Myr at $z=6$.  A steep decrease in detection sensitivity occurs at redshift $z>10$, but is is relatively constant at $10^8$\Msun\ for $z<10$, so we adopt constant detectivity in mass.  (The gap at $z\sim6.5$ is an artifact of the way redshifts are determined by interpolation in a scheme that is working near its resolution limit---see \citealt{2024ApJ...964..150D} for details.) This figure is reproduced with permission from \cite{2024ApJ...964..150D}.}  

\label{SFH-1Gyr}
\end{figure*}

This paper explores an alternative route, one that we believe is more illuminating than attempting to reproduce the growing starlight in the earliest times.  Numerical simulations are fundamentally based on \emph{mass}, with light---subject to significant uncertainty---being a secondary consideration. Fortunately, observations can be made that measure the stellar mass in these early galaxies, allowing for a direct comparison of stellar mass growth for the earliest galaxies (representing the production of chemical enrichment that is crucial) with the production of stellar mass that theoretical simulations track.

\defcitealias{2024ApJ...964..150D}{AD24}\citet[][hereafter \citetalias{2024ApJ...964..150D}]{2024ApJ...964..150D}, described the fortunate correspondence of the timescale of hundreds of millions of years for this critical phase of galaxy growth with the lifetimes of A-stars.  Assuming that the stellar mass function has not significantly evolved from what it was when this all began to what it is today---as observations support \citep{2014ApJ...796...75C} the light detected from imaging these early galaxies in multiple near-infrared bands is directly measuring $\gappr90\%$ of the stellar mass. The goal of this study, then, is to compare the observations made by \citetalias{2024ApJ...964..150D} with the predictions of a numerical simulation of galaxy birth and growth---{\sc Galacticus} \citep{2012NewA...17..175B}.

\section{Methods}

\subsection{Early Star Formation Histories of Galaxies Observed by JWST}

\emph{JWST} observations have triggered a deluge of theoretical work that focuses on and examines the population of luminous galaxies at high redshifts. Early studies (e.g., \citealt{2023NatAs...7..731B}) suggested that there was a tension with theory, as even 100\% efficiency of conversion of baryons into stars could not produce a sufficient number of such bright galaxies. As observational analyses have improved (and more careful statistical analyses have been performed, see, for example \citealt{2024arXiv241011680D,2025ApJ...982...23J}), this tension has weakened, but it is still challenging to make enough high-luminosity galaxies in state-of-the-art simulations without invoking significant modifications to the initial mass function \citep[][but see \citealt{2025MNRAS.536..988F} for a counterexample]{2025MNRAS.536.1018L,2025ApJ...980...10J,2025arXiv250418618Y}.

Focusing instead on the growth of stellar mass offers a cleaner approach---one that isolates star formation (mass) from stellar evolution, which involves many `variables,' for example, elemental yields within and between generations, obscuration by dust, and contamination by AGN light. In galaxies with ages less than $\sim$1~Gyr, the light from A stars dominates: lower mass stars have been born, of course, but their light is overwhelmed by that of A stars (essentially the same effect as encountered when studying a young globular cluster). More massive O \& B stars, though much more luminous, have short lifetimes: any stellar population older than 100~Myr has very few O \& B stars compared to A-stars, and they contribute far less to the total mass. By happenstance, A-star spectra are essentially free of metal absorption lines, so heavy-element abundances do not affect the emitted starlight---a huge advantage. In summary, to the extent that the initial mass function (IMF) of star birth is not changing rapidly over this period, the amount of light emitted by A-stars dominates, yielding a direct measure of the mass of a stellar system. Furthermore---and the key point---the fact that A-star spectra from A0 to A9 are easily distinguishable means that a non-negative least-squares (NNLS) solution for a galaxy with $12<z<6$ yields a set of vector spectra that reveal the SFH of the galaxy as stellar mass in each epoch, summing to its total stellar mass at observation (see \citealt{2023ApJ...947...L27}).

Figure~\ref{SFH-1Gyr}, drawn from \citetalias{2024ApJ...964..150D} Figure 6, plots the growth of stellar mass in discrete stellar systems over the interval $6\le z \le12$ for a sample of $\sim$1000 young galaxies over a field of $\sim$25 sq.~arcmin.  The time interval is divided into integer `epochs' containing galaxies whose redshifts are known to a precision of $\sim$1\%. This is the quantity that we want to compare to simulations---stellar mass growth, rather than the evolution of galaxy brightness.

\citetalias{2024ApJ...964..150D} used \emph{JWST} photometry in 7 near-infrared bands to define four types of star formation histories (SFHs) that distinguish bursts from rates that slowly evolve over several epochs. Figure 1 of \citetalias{2024ApJ...964..150D} shows examples of 4 different SFHs: \emph{burst} (one epoch), \emph{contiguous} (two bursts close in time), \emph{stochastic} (multiple discrete bursts), and \emph{continuous} over 4 epochs---categories were defined by epoch, not mass.  The output of the SEDz code used for this analysis is the stellar mass produced over 7 epochs ($z=6, 7 \ldots 12$) for each galaxy, and a burst is defined as a galaxy for which there is detected star formation in 1 or 2 adjacent epochs and no stellar mass detected in the other epochs.  A dot in Figure~6 of \citetalias{2024ApJ...964..150D} (Figure~1 in this paper) shows its \emph{total mass} and its color encodes the SFH type, for which time---not mass---is a parameter.

Our use of \textsc{Galacticus} to compare observed SFHs to theory requires a different burst definition. For this, we chose to compare \emph{total star formation} over $5.75<z<7.75$ (redshift `bins' 6 and 7) to that in $7.75<z<12.75$ (redshift bins 8, 9, 10, 11, 12). Because the \citetalias{2024ApJ...964..150D} data become noticeably incomplete below $10^8\Msun$, we chose a detection threshold of $\ge2\times10^8\Msun$---for bins (6+7) and $<10^8\Msun$ for (8+9+10+11+12).  Thus, the `burst' definition in this paper will allow us to explore detail evolution from \textsc{Galacticus} SFHs for galaxies that meet SFH type 1 in \citetalias{2024ApJ...964..150D}.

By necessity, star formation histories derived this way---from broad-band photometry---have poor time resolution, so the present work began as an attempt to use theoretical simulations with high-time-resolution ($\sim$10~Myr) to see which time-resolved histories might match \citetalias{2024ApJ...964..150D} `burst' vs. `non-burst' categories. We were well along in pursuit of this goal before we realized that matching the \emph{overall} 600 Myr growth of stellar mass for data and model was a prerequisite for this task. As we describe below, we were surprised to discover that in this basic property, the semi-analytic model {\sc{Galacticus}} was in good agreement ``\emph{ab initio}"---without `fine tuning' of parameters. {\bf This is the main result of this paper}.  The confidence gained from this discovery encouraged our use of {\sc{Galacticus}} as a tool in understanding the various histories revealed in the \emph{JWST} data.

We evaluate how well a simulation matches the \emph{JWST} data in four ways. First, as just described, we can compare the integral of stellar mass over redshift of the sample in Figure~\ref{SFH-1Gyr}---from $z=12$ to $z=6$. (Reconciling stellar mass growth for data and model over this whole period is a prerequisite.) Second, we can identify bursts in the simulation by applying a definition of `bursts' with higher time-resolution that is consistent with the fraction found in the \citetalias{2024ApJ...964..150D} `epoch-based' definition, where a frequency of $\sim$50\% was measured. Third, we can compare `burst'  and `non-burst' histories to see, for example, whether mergers play a similar or different role, as might be the case if mergers are gas-rich as opposed to gas-poor, or if mass growth is dominated by accretion. Fourth, we can compare the SFHs of `bursts` and `non-bursts' in contributing to the mass buildup and look for consistency in specific properties. For example, if the simulation finds that mergers are a significant component in `bursts,' is this consistent or inconsistent with the data in Figure~\ref{SFH-1Gyr}?

In the next section, we describe how we measured these behaviors in the theoretical model.

\subsection{Early SFHs of Galaxies in {\sc Galacticus}}\label{sec:methodsModel}

%\textbf{Describe the modeling that we do in %Galacticus, highlighting the physics included, %particularly that driving star formation and %bursts.}

We use the {\sc Galacticus} semi-analytic model\footnote{Specifically revision \href{https://github.com/galacticusorg/galacticus/commit/35bbde0a9589995f07440bc0a616b6005a1c3111}{35bbde0}.} \citep{2012NewA...17..175B} to compute the properties of galaxies at $z=6$. Specifically, we use the implementation of the model detailed in \citeauthor{2018MNRAS.474.5206K}~(\citeyear{2018MNRAS.474.5206K}; \S2.2) and most recently described by \cite{2024arXiv241011680D}. This model was calibrated to approximately match a set of low redshift data, including galaxy stellar mass functions.

We refer the reader to the above works for full details of the model, but describe the model for feedback here, as this is the only aspect of the model that we vary in this work. In the {\sc Galacticus} model, feedback due to star formation is implemented as an outflow rate:
\begin{equation}
\dot{M}_\mathrm{outflow} = \left(\frac{V_\mathrm{outflow}}{V_\mathrm{galaxy}}\right)^{\alpha_\mathrm{outflow}} \Psi,
\end{equation}
where $V_\mathrm{galaxy}$ is the rotation curve of the galaxy measured at the scale radius of the disk or spheroid component, $\Psi$ is the star formation rate, $V_\mathrm{outflow}$ is a characteristic velocity above which outflows become negligible, and $\alpha_\mathrm{outflow}$ determines how the strength of outflows varies with galaxy mass. In \cite{2018MNRAS.474.5206K} values of $V_\mathrm{outflow}=250$~km/s and $\alpha_\mathrm{outflow}=3.5$ were used. \cite{2024arXiv241011680D} constrained these parameters to match data from \emph{JWST} CEERS and NGDEEP surveys, finding 95\% confidence intervals of $V_\mathrm{outflow}=150^{+280}_{-60}$~km/s and  $\alpha_\mathrm{outflow}=1.5^{+0.5}_{-0.9}$. The specific feedback model and parameter values strongly influence model results at high redshifts, as shown by \cite{2024arXiv241011680D}, and as we will show in this work. Therefore, we will allow these parameters to vary in this work to examine how feedback influences stellar mass growth at high redshifts. We begin with a ``default'' choice of parameters of $V_\mathrm{outflow}=150$~km/s (in both disk and spheroid components of model galaxies) and $\alpha_\mathrm{outflow,(disk|spheroid)}=(3.8,2.7)$---chosen to provide reasonable matches to low redshift stellar mass functions.

To classify model galaxies, we follow a procedure mimicking the observational analysis of \citetalias{2024ApJ...964..150D}. Specifically, for each galaxy identified at $z=6$, we determine the mass of stars that it formed in the redshift intervals, 6--7, 7--8, etc., labeling these $M_{\star,6-7}, M_{\star,7-8}$, etc. Any galaxy for which $M_{\star,6-7}+M_{\star,7-8} > 2\times10^8\mathrm{M}_\odot$ is considered to be observationally detectable. We subdivide those galaxies into `burst' and `non-burst' subsamples based on the maximum mass of stars formed across all earlier redshift intervals, $M_{\star,\mathrm{max}} = \hbox{max}(M_{\star,8-9}, M_{\star,9-10},\ldots)$. If $M_{\star,\mathrm{max}} < 10^8\mathrm{M}_\odot$ (i.e. the galaxy formed less than $10^8\mathrm{M}_\odot$ of stars in any earlier epoch) the galaxy is considered to be a `burst' at $z=6$---it has a substantial amount of recent star formation, together with little star formation at earlier times. Alternatively, if $M_{\star,\mathrm{max}} \ge 10^8\mathrm{M}_\odot$ the galaxy is considered to be a `non-burst'--the rate of recent star formation is not substantially greater than that at earlier times.

\section{Results}

\begin{figure*}
  \includegraphics{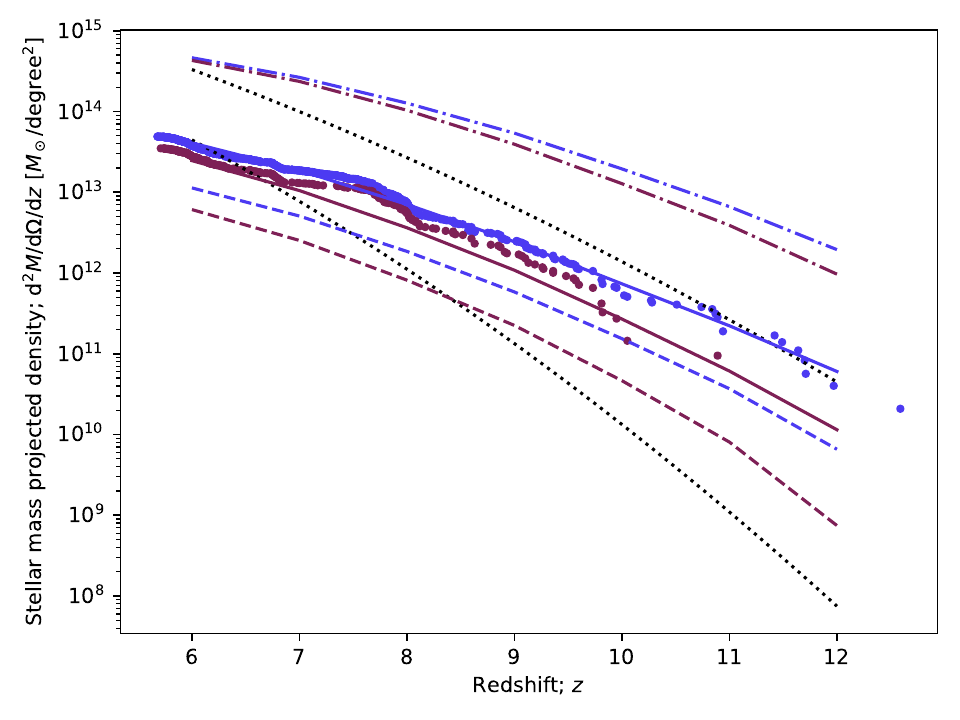}
 \caption{The growth of stellar mass (shown as the total stellar mass in galaxies per square degree per unit redshift) from $z=12$ to $z=6$. Colors indicate the stellar mass selection: blue shows the total mass in galaxies with stellar masses above $10^{7.8}\mathrm{M}_\odot$ while purple shows the total mass in galaxies with stellar masses above $10^{8.6}\mathrm{M}_\odot$. Circles indicate observed galaxies from \citetalias{2024ApJ...964..150D}. Lines show results from our {\sc Galacticus} model, with dashed lines showing our default feedback model ($V_\mathrm{outflow}=150$~km/s; $\alpha_\mathrm{outflow,(disk|spheroid)}=(3.8,2.7)$), and solid lines indicating our reduced feedback model ($V_\mathrm{outflow}=120$~km/s; $\alpha_\mathrm{outflow,(disk|spheroid)}=(3.0,3.0)$). For comparison, we also show results from a very weak feedback model ($V_\mathrm{outflow}=20$~km/s; $\alpha_\mathrm{outflow,(disk|spheroid)}=(2.0,2.0)$) as dot-dashed lines. Black, dotted lines show the total baryonic mass (assuming the universal baryon fraction) in halos with masses above $10^{10.7} \mathrm{M}_\odot$ and  $10^{11.5} \mathrm{M}_\odot$ (upper and lower lines, respectively).}
 \label{fig:stellarMass}
\end{figure*}

Figure~\ref{fig:stellarMass} shows the growth of the total stellar mass in galaxies above stellar mass thresholds of $10^{7.8}~\mathrm{M}_\odot$ (blue) and $10^{8.6}~\mathrm{M}_\odot$ (purple) per square degree, per unit redshift from $z=12$ to $z=6$. Circles indicate observational measurements from \citetalias{2024ApJ...964..150D}, while lines indicate results from our {\sc Galacticus} models. Our default feedback model, chosen to produce a reasonable match to galaxy stellar mass functions at $z=0$, is shown by the dashed lines. While this model exhibits a similar trend with redshift as the observations, it is offset lower in total stellar mass at all redshifts by a factor of at least 4. To assess the sensitivity of this offset to choices of model parameters, we also show, as solid lines, a {\sc Galacticus} model in which the feedback parameters are adjusted to $V_\mathrm{outflow}=120$~km/s, $\alpha_\mathrm{outflow}=3.0$ to weaken feedback (particularly in lower mass galaxies). This weaker feedback model matches the observed data quite well across the entire redshift range, indicating that the assembly of stellar mass at these high redshifts is very sensitive to the strength of feedback. For comparison, we also show, using dot-dashed lines, a model with very weak feedback ($V_\mathrm{outflow}=20$~km/s, $\alpha_\mathrm{outflow}=2.0$) which hugely overpredicts the amount of stellar mass formed at all redshifts. However, we note that the trend of stellar mass growth with redshift is very similar in both feedback models. Black, dotted lines in Figure~\ref{fig:stellarMass} show the total baryonic mass (assuming the universal baryon fraction) in halos with masses above $10^{10.7} \mathrm{M}_\odot$ and  $10^{11.5} \mathrm{M}_\odot$ (upper and lower lines respectively), chosen to approximately match the observed data at $z=12$ and $z=6$ respectively. Neither of these curves matches the redshift dependence seen in the observed data. This clearly shows that the star formation efficiency (the fraction of available baryons turned into stars) \emph{must} be redshift-dependent---either by varying with redshift at fixed halo mass, or by having a redshift-dependent halo mass threshold above which star formation ``switches on'' (and below which the star formation efficiency is negligible).

For comparison, we have verified that our model agrees quite closely with results from the FIREBox-HR simulations of \cite{2023MNRAS.522.3831F}, which predict, at $z=6.33$, a stellar mass density of $\mathrm{d}^2 M / \mathrm{d}z / \mathrm{d}\Omega = 2.27 \times 10^{13} \mathrm{M}_\odot \mathrm{deg}^{-2}$ for galaxies with $M_\star > 10^{7.8} \mathrm{M}_\odot$ (R. Feldman, private communication).

\setlength{\tabcolsep}{1pt}

\begin{figure*}
 \begin{tabular}{ccc}
  \includegraphics[width=57mm]{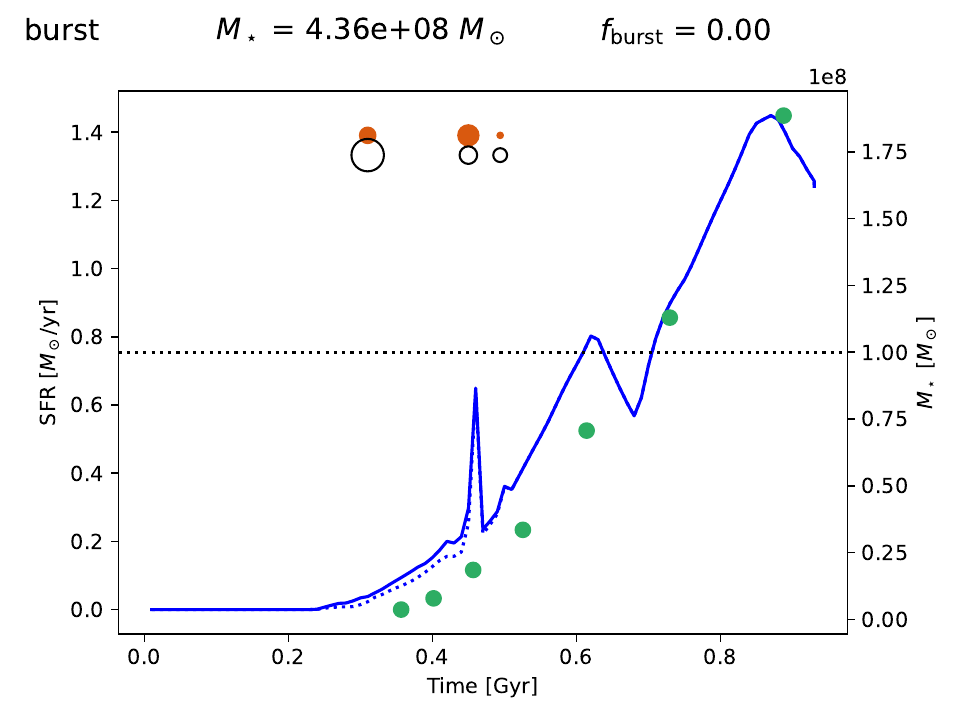} &
  \includegraphics[width=57mm]{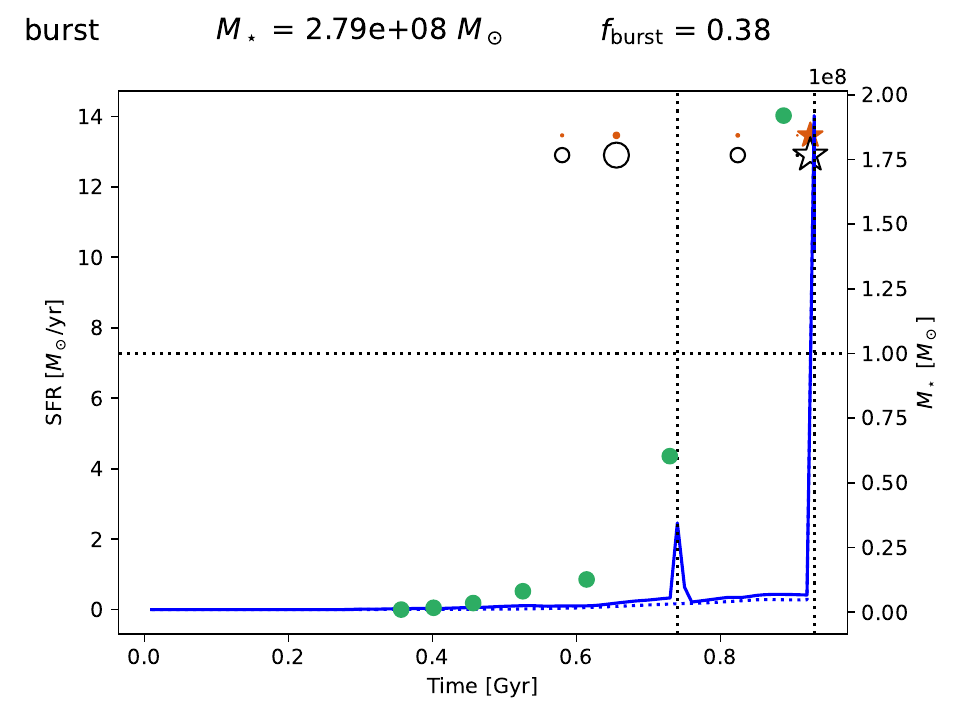} & 
  \includegraphics[width=57mm]{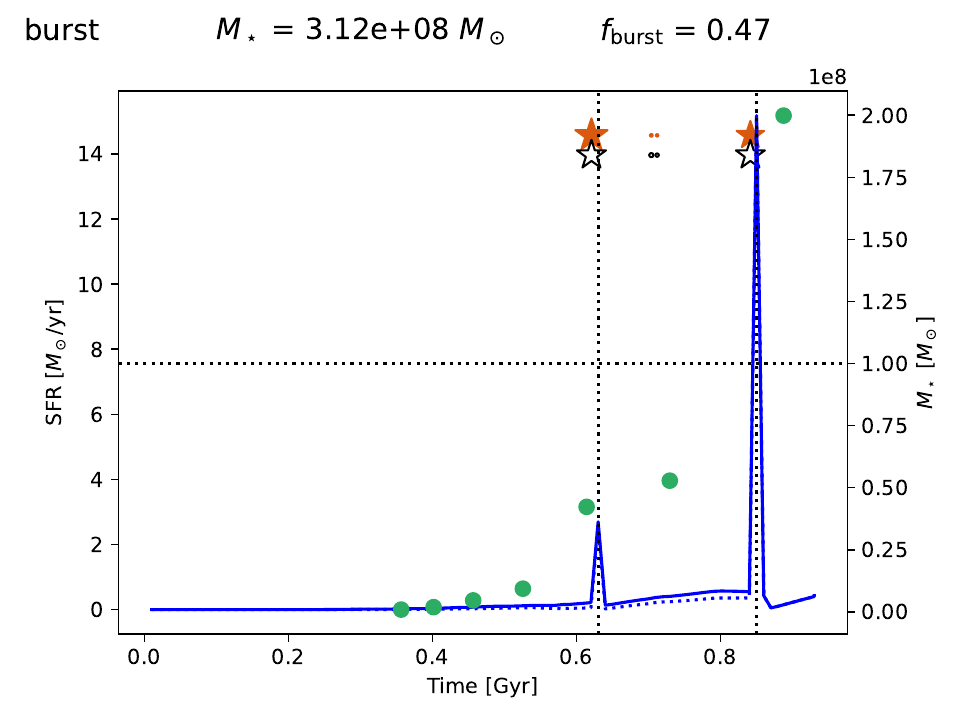} \\
  \includegraphics[width=57mm]{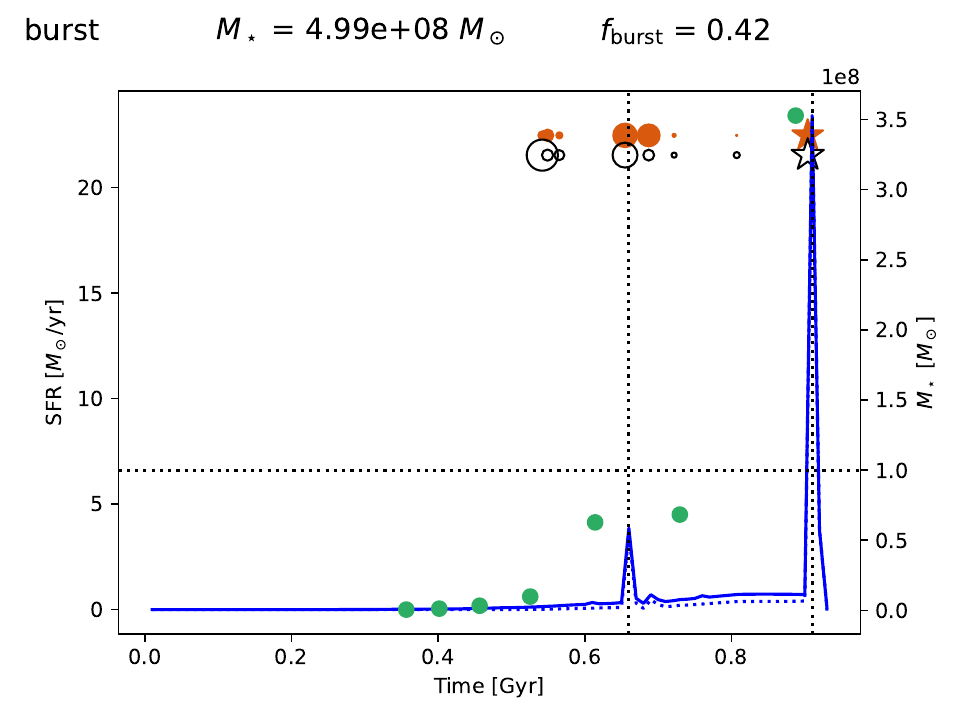} &
  \includegraphics[width=57mm]{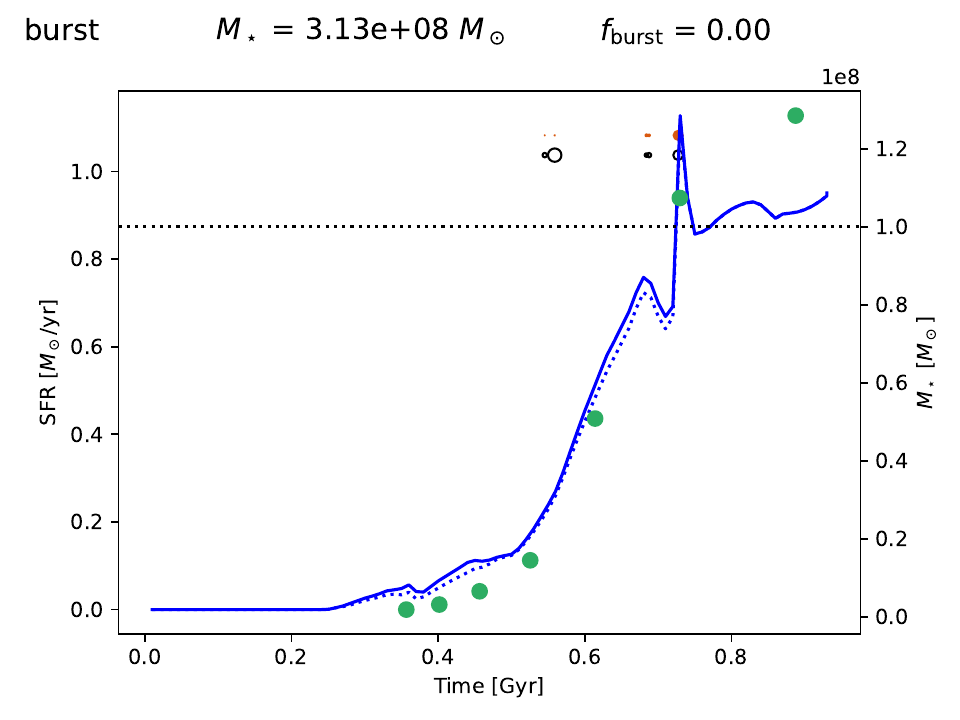} & 
  \includegraphics[width=57mm]{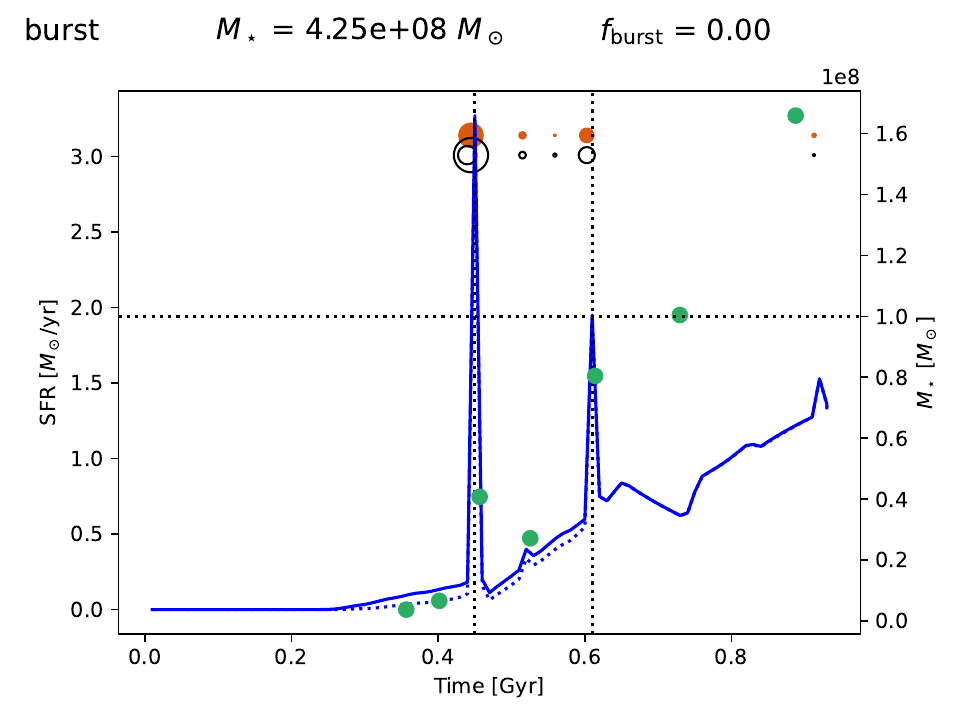} \\
  \includegraphics[width=57mm]{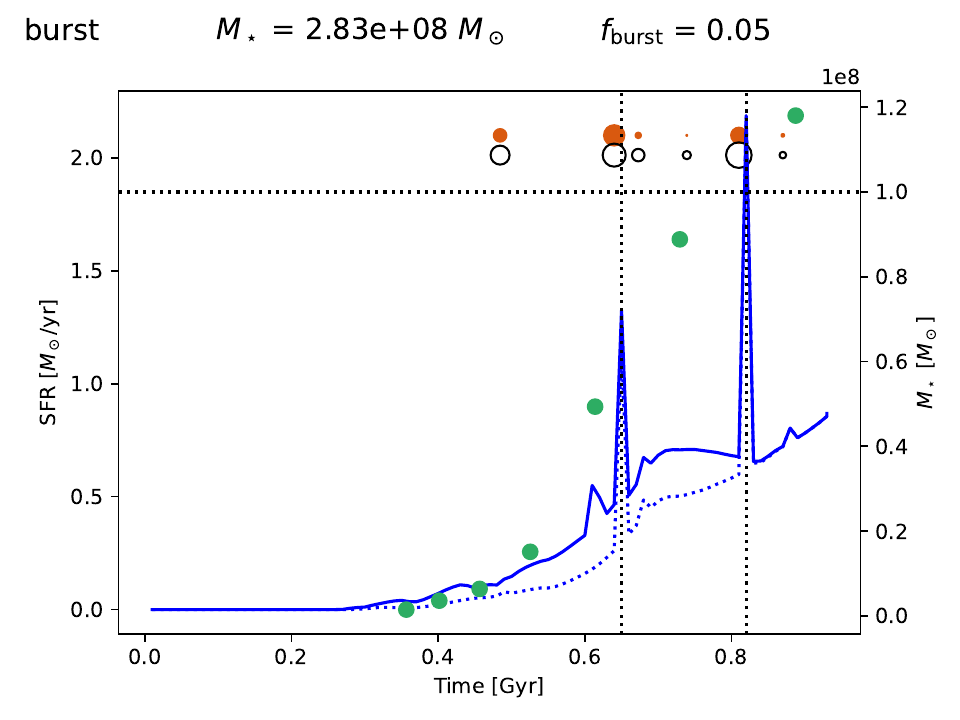} &
  \includegraphics[width=57mm]{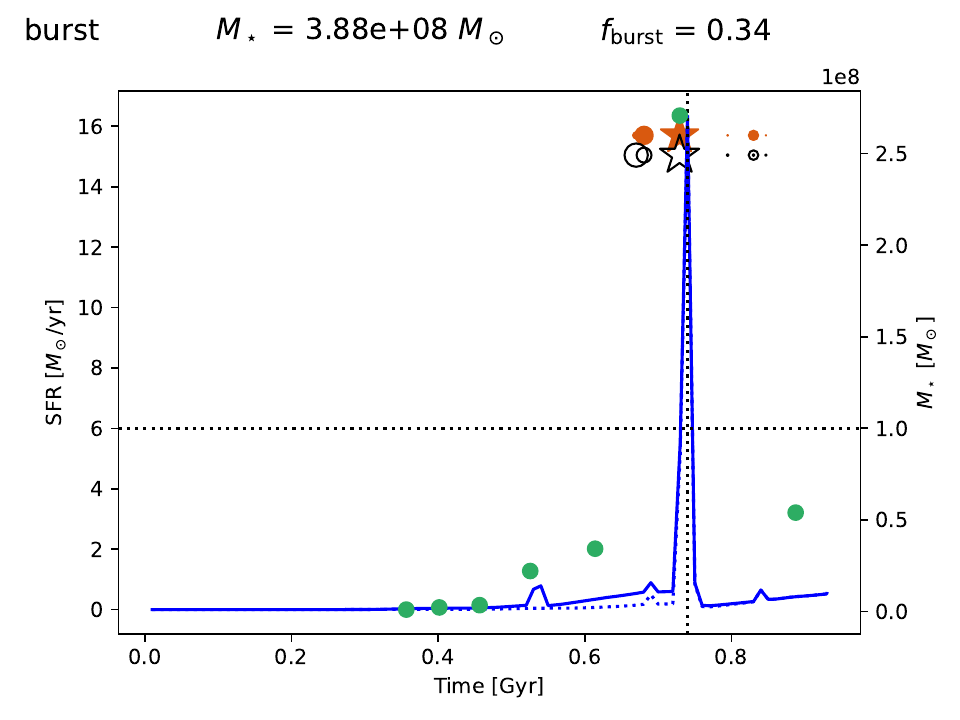} & 
  \includegraphics[width=57mm]{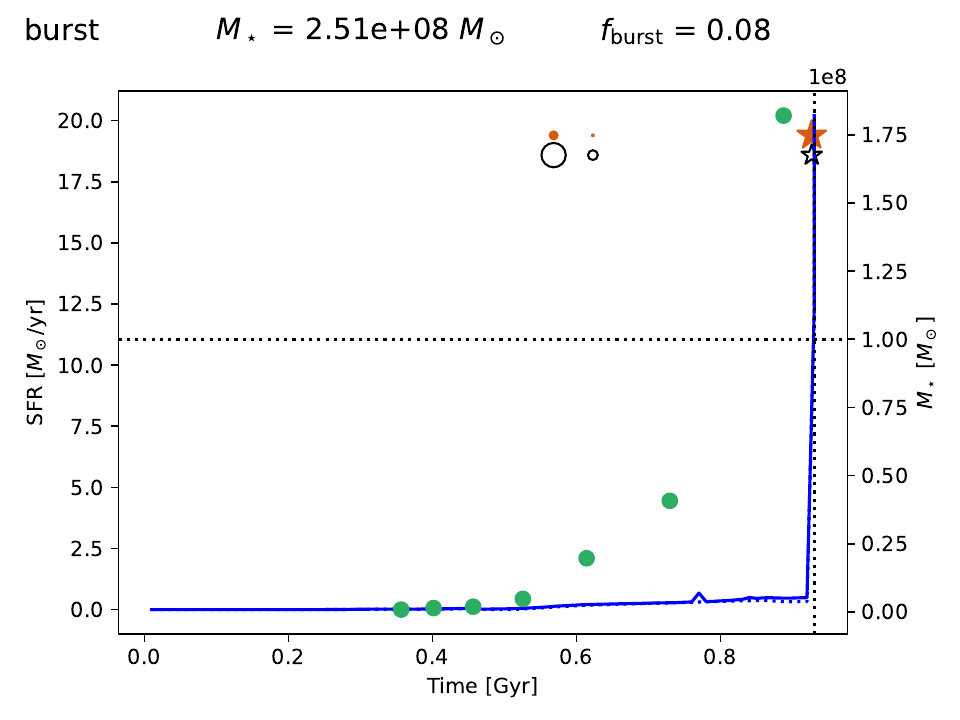} \\
 \end{tabular}
 \caption{Examples of star formation histories of {\sc Galacticus} central galaxies classified into the `burst' category. The total stellar mass at $z=6$ is shown above each panel, along with the fraction of stellar mass formed in `starbursts' (indicated by the vertical dotted lines, and identified as times at which the star formation rate in a peak exceeds twice the mean star formation rate) in the $z=7$ and $z=6$ bins. Solid blue lines show the total star formation rate (left axis), while dotted blue lines indicate the \emph{in situ} star formation rate. Green circles show the integrated stellar mass (right axis) formed in each unit redshift interval. The horizontal dotted line indicates a stellar mass of $10^8\mathrm{M}_\odot$ as used in the classification of galaxies into `bursts' and `non-bursts.' Open black circles/stars (shown at arbitrary position on the $y$-axis) indicate the stellar mass ratio (defined as the stellar mass of the satellite galaxy divided by the sum of the stellar masses of the satellite and central galaxies) in merger events, while filled red circles/stars show the gas mass ratio in the same merger events---larger circles indicate larger mass ratio. Black/red circles indicate minor mergers, while black/red stars indicate major mergers.}
 \label{fig:bursts}
\end{figure*}

\begin{figure*}
 \begin{tabular}{ccc}
  \includegraphics[width=57mm]{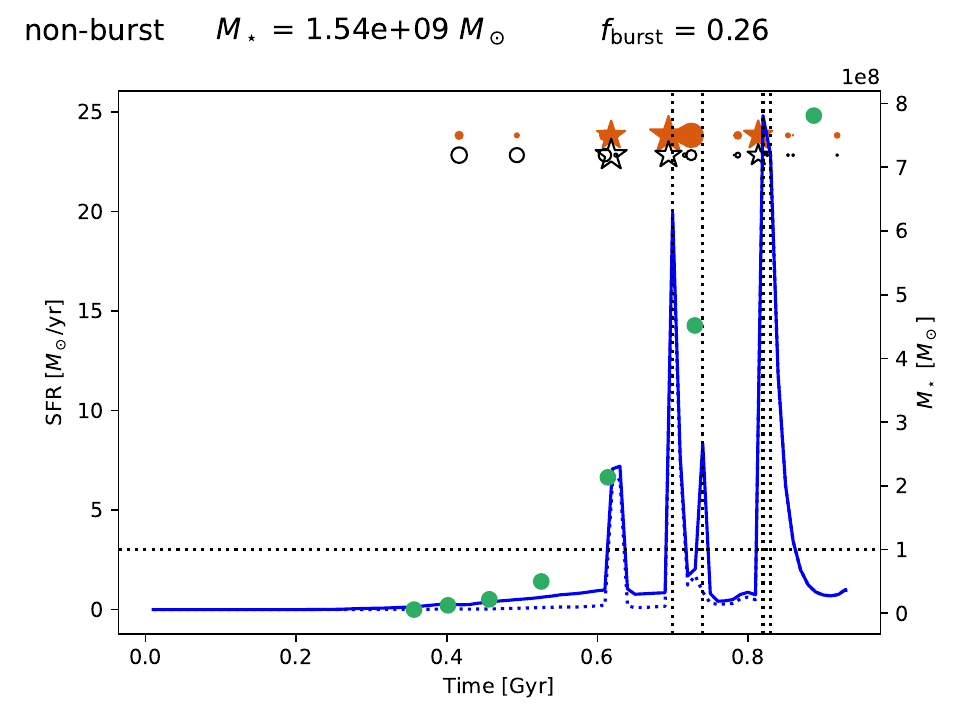} &
  \includegraphics[width=57mm]{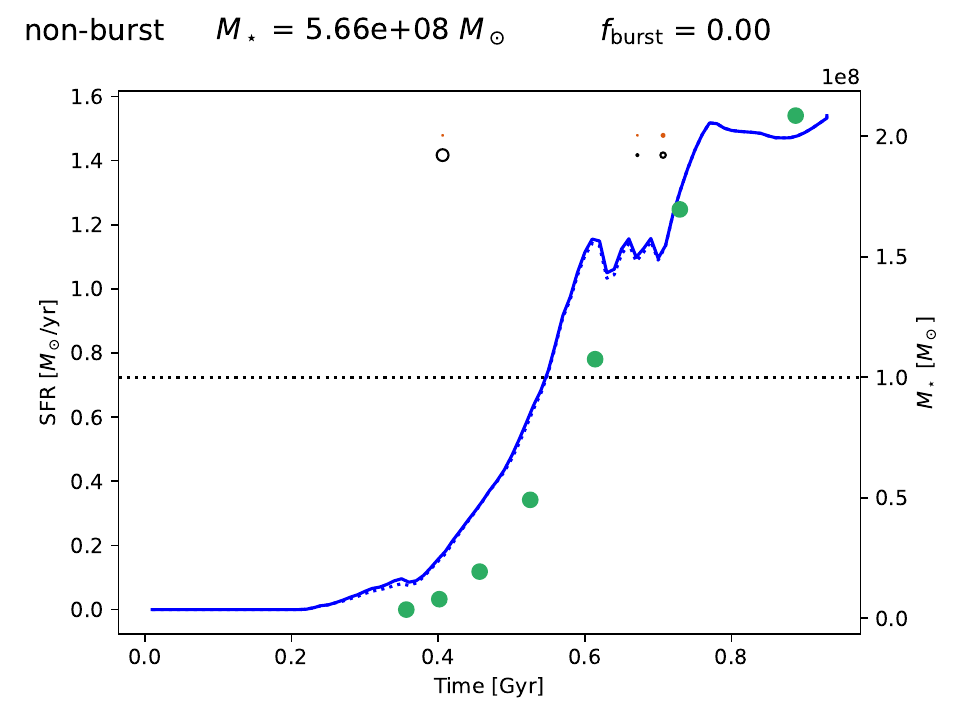} & 
  \includegraphics[width=57mm]{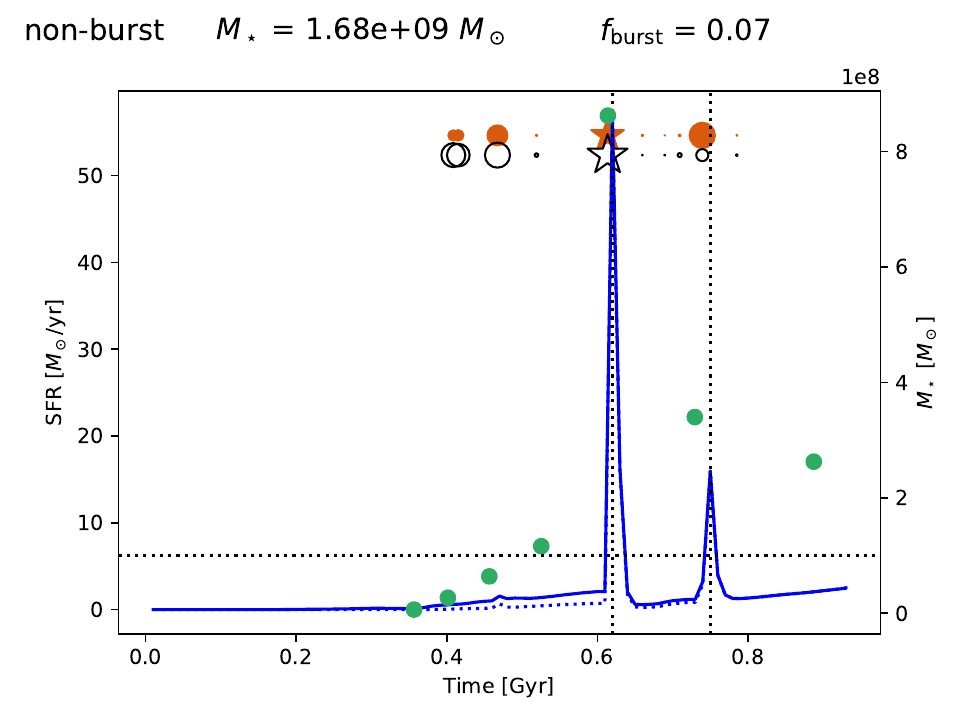} \\
  \includegraphics[width=57mm]{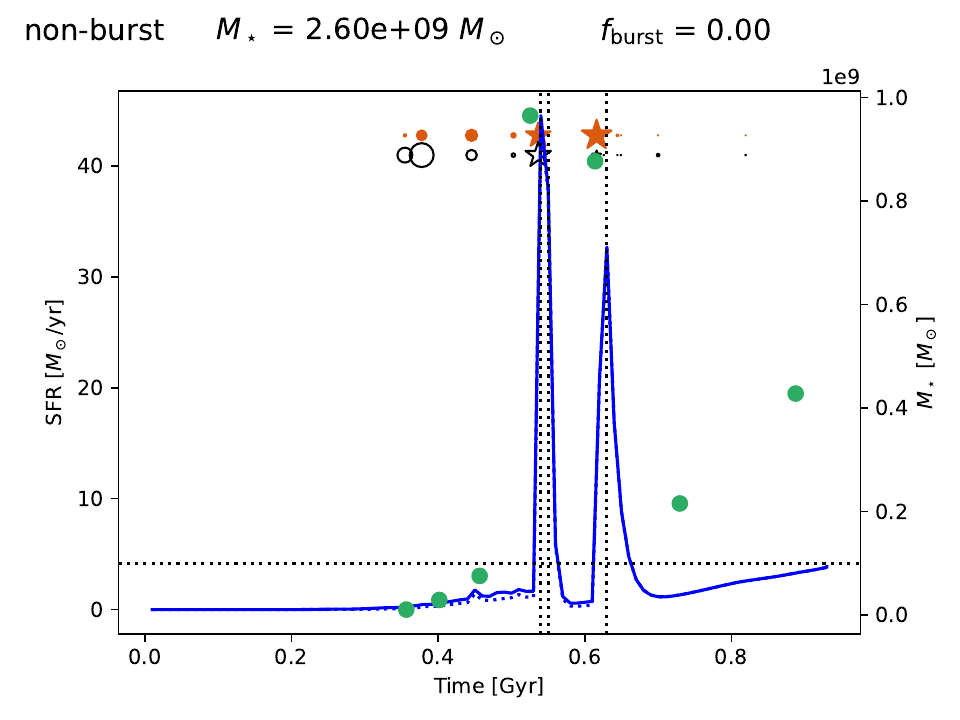} &
  \includegraphics[width=57mm]{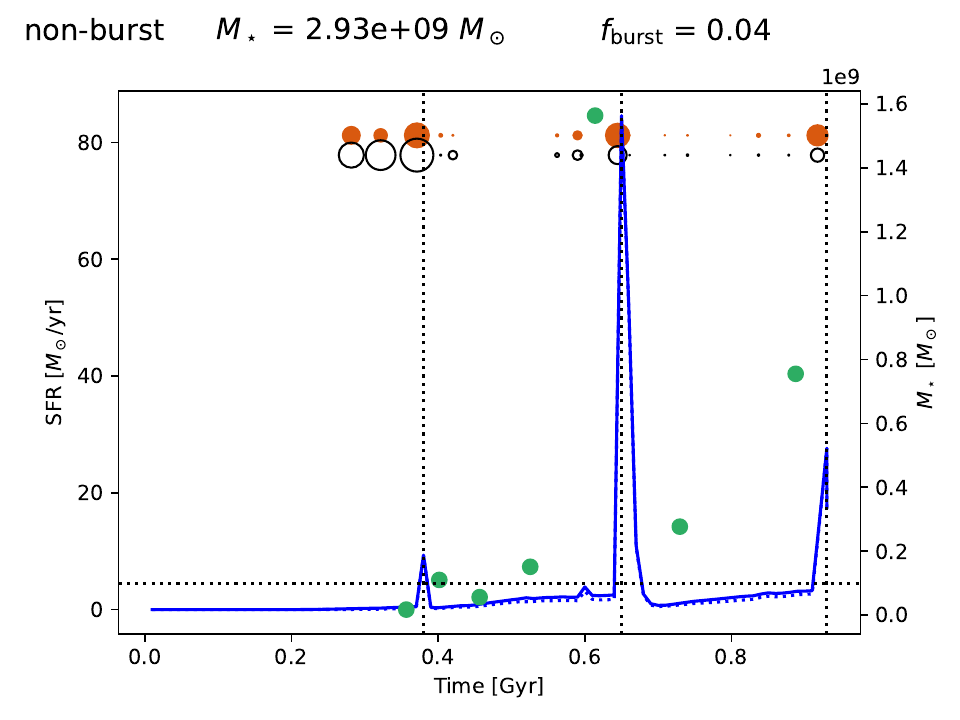} & 
  \includegraphics[width=57mm]{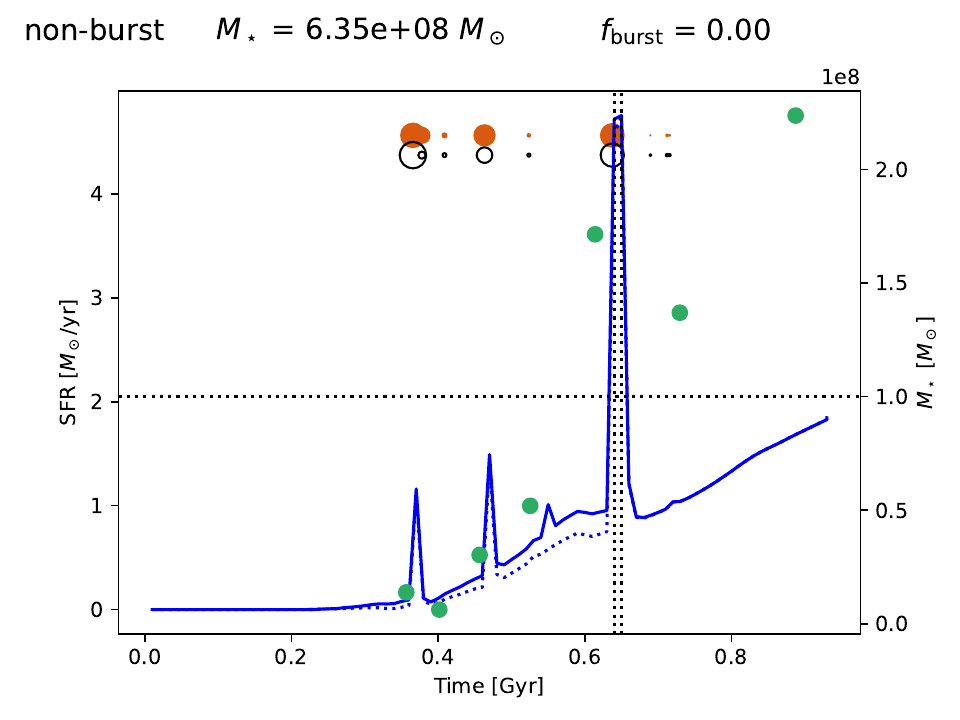} \\
  \includegraphics[width=57mm]{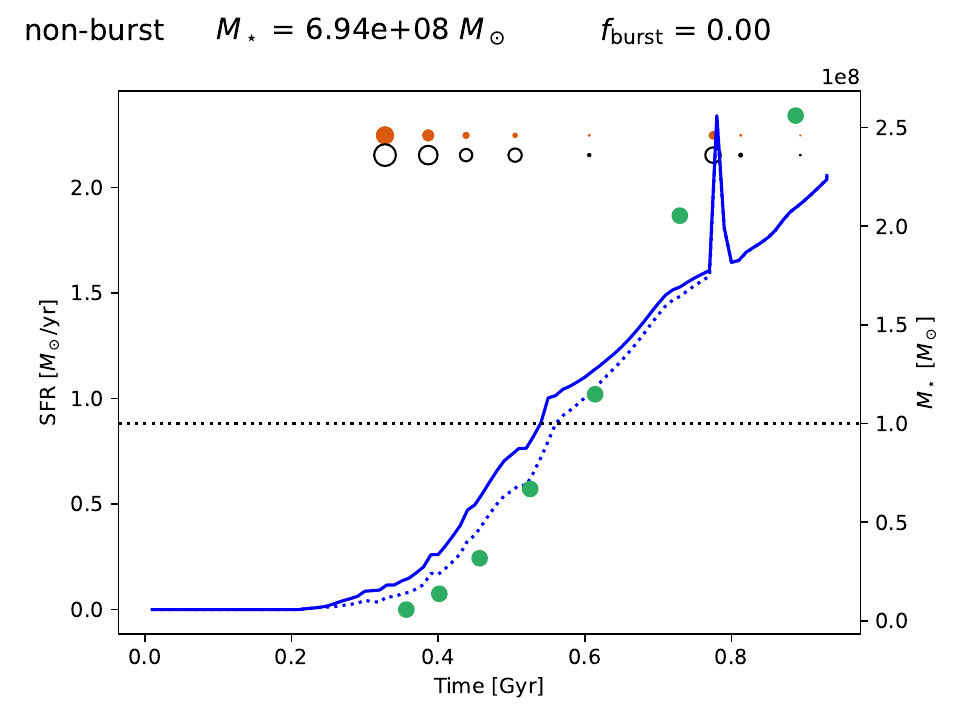} &
  \includegraphics[width=57mm]{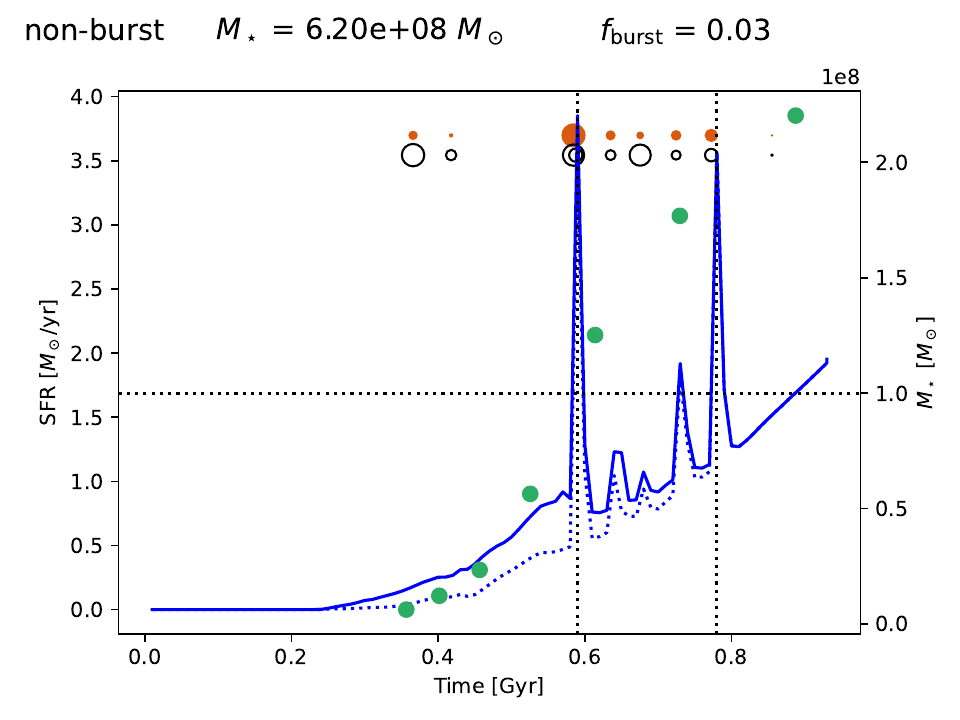} & 
  \includegraphics[width=57mm]{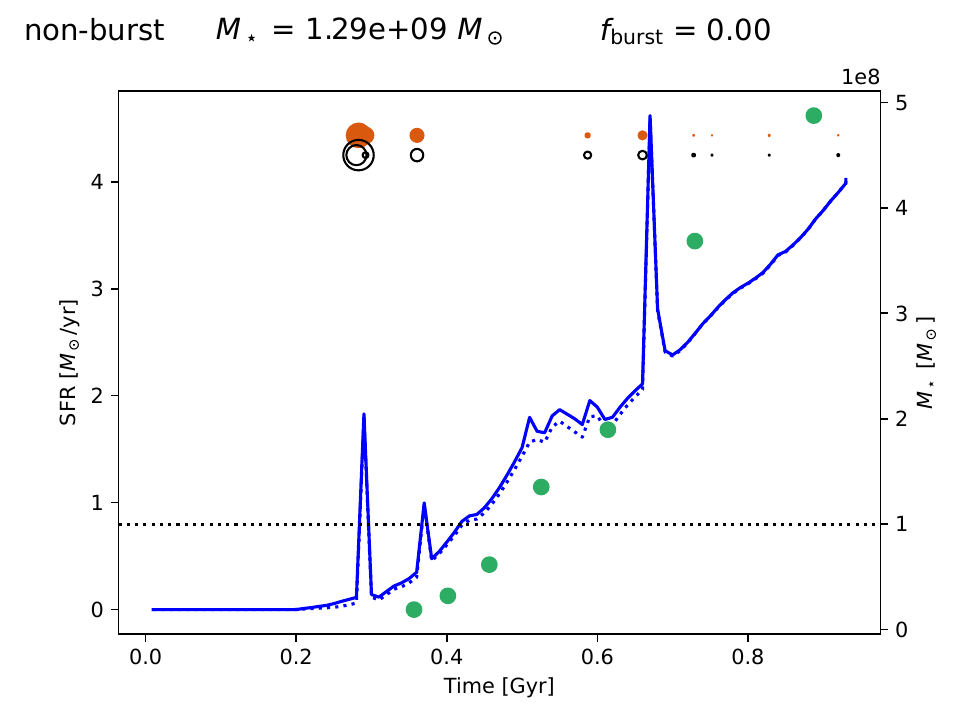} \\
 \end{tabular}
 \caption{Examples of star formation histories of {\sc Galacticus} central galaxies classified into the `non-burst' category.  The total stellar mass at $z=6$ is shown above each panel, along with the fraction of stellar mass formed in bursts (indicated by the vertical dotted lines, and identified as times at which the star formation rate in a peak exceeds twice the mean star formation rate) in the $z=7$ and $z=6$ bins. Solid blue lines show the total star formation rate (left axis), while dotted blue lines indicate the \emph{in situ} star formation rate. Green circles show the integrated stellar mass (right axis) formed in each unit redshift interval. The horizontal dotted line indicates a stellar mass of $10^8\mathrm{M}_\odot$ as used in the classification of galaxies into `bursts' and `non-bursts.' Open black circles/stars (shown at arbitrary position on the $y$-axis) indicate the stellar mass ratio (defined as the stellar mass of the satellite galaxy divide by the sum of the stellar masses of the satellite and central galaxies) in merger events, while filled red circles/stars show the gas mass ratio in the same merger events---larger circles indicate larger mass ratio. Black/red circles indicate minor mergers, while black/red stars indicate major mergers.}
 \label{fig:smooths}
\end{figure*}

Following the classification of model galaxies into `burst' and `non-burst' classes (as described in \S\ref{sec:methodsModel}), we find that our weaker feedback {\sc Galacticus} model predicts that 67\% of galaxies at $z=6$ fall into our `burst' category, compared to 53\% from the observational sample of \citetalias{2024ApJ...964..150D}. `Burst' galaxies have lower mean stellar mass ($\log_{10} M_\star/\mathrm{M}_\odot=8.4$) at $z=6$ compared to their `non-burst' counterparts ($\log_{10} M_\star/\mathrm{M}_\odot=9.0$), and have experienced a approximately the same mean number of major mergers---0.46 per galaxy, compared to 0.40 per galaxy for their non-burst counterparts. As can be seen from this number, not all burst galaxies have experienced a major merger in their history.

Figures~\ref{fig:bursts} and \ref{fig:smooths} show Galacticus star formation histories are parsed into the `burst' and `non-burst' categories, respectively. Here, we select only central galaxies\footnote{In practice, the contribution of satellite galaxies to this sample would be entirely negligible.}, and we remind the reader that this classification is made based on the integrated stellar mass formed in unit redshift intervals. Galaxies that formed less than $10^8\mathrm{M}_\odot$ in any redshift interval prior to $z=7$ but formed more than $2 \times  10^8\mathrm{M}_\odot$ in the combined $z=7$ and $z=6$ intervals are considered to be `bursts'. The green circles in figures~\ref{fig:bursts} and \ref{fig:smooths} show the integrated stellar mass formed in each unit interval (with the scale shown on the right axis), with the horizontal dotted line indicating the $10^8\mathrm{M}_\odot$ threshold.

The stellar mass buildup (Fig. 2) and `burst fraction' (67\% compared to 53\%) are essentially the same for observations and model. It is therefore sensible to infer something else---the presence of shorter timescale enhancements in star formation of duration 30--40 Myr in the simulations that cannot be resolved in the observational data, and which we will call `starbursts'.  We associate these jumps in star-formation rate with those observed in modern galaxies and associated with giant HII regions, or star-formation surges at the galaxy-scale seen in galaxy mergers.  We will use this additional information to infer if the observed \emph{`bursts'} are made up of canonical \emph{starbursts} or whether they measure increasing rates of star formation on a galactic scale at these high redshifts.

The vertical dotted lines in Figures~\ref{fig:bursts} and \ref{fig:smooths} mark the times of starbursts in the model star formation histories (shown by solid blue lines, with the scale shown on the left axis). To identify `starbursts' in the full model star formation histories, we first identify peaks and compute the excess star formation rate in each peak above that expected by linearly interpolating the star formation rate between the two adjacent snapshots---this gives an estimate of the increase in star formation associated with this peak of star formation. A peak is identified as a `starburst' if this excess star formation rate exceeds twice the mean star formation rate in the galaxy (averaged over the entire history up to $z=6$).

In these figures, we show the stellar mass ratio (defined as the stellar mass of the satellite
galaxy divided by the sum of the stellar masses of the satellite and central galaxies) in merger events as open black circles/stars (at arbitrary position on the $y$-axis), and the gas mass ratio in the same merger events as filled red circles/stars, with larger circles/stars indicating larger mass ratio. Minor mergers are indicated by circle symbols, while major mergers are indicated by star symbols.

We begin our discussion of Figures~\ref{fig:bursts} and \ref{fig:smooths} by comparing the contribution of `starbursts' in the low-time-resolution `burst' and `non-burst' histories.
The top-center panel of Figure~\ref{fig:bursts} shows a case where the classification of a galaxy as a \emph{burst} aligns well with the more time-resolved model star formation history. In this galaxy, the integrated stellar mass in each unit redshift interval is gradually increasing with time, but shows starbursts at redshifts 6 and 7, causing the galaxy to be placed into the `burst' category. In particular, the jump in integrated stellar mass coincides closely with a starburst in which the star formation rate briefly reaches around 14 $\mathrm{M}_\odot$/yr. In this it is clear that a large mass ratio, gas-rich merger occurs and triggers the starburst (see Fig. 3 caption).

In contrast, the top-left panel of Figure~\ref{fig:bursts} is a growth history without `starbursts' in redshifts 6 \& 7.  Its integrated stellar mass formed in unit redshift intervals increases monotonically, in such a way that the integrated stellar masses remain below the detection threshold of $10^8\mathrm{M}_\odot$ prior to the $z=7$ bin, but exceed it at later times---our definition of a `burst.' However, there are no `starbursts` using the simulation-motivated definition, just two modest enhancements of star formation rate associated with halo growth or minor galaxy mergers. In the remaining panels of Figure~\ref{fig:bursts}---center-center and center-right, and bottom-left--are further examples of galaxies that are `starburst'-free (following the definition given above, which requires that the excess star formation rate in any peak in the star formation history exceed twice the average star formation rate of the galaxy) but in the `burst' category. We find that 56\% of galaxies classified into the `burst' category have a \emph{starburst} (following our definition based on the theoretical star formation histories) in the $z=7$ or $z=6$ bins.

Figure~\ref{fig:smooths} similarly shows star formation histories for galaxies classified into the `non-burst' category.  Here, the top-center panel shows a galaxy with a relatively smooth growth history, with the integrated stellar mass in each unit redshift interval increasing monotonically and just a few small peaks---insufficiently high to be labeled as `starbursts' in our model.  In contrast, the top-left panel shows a case of steady monotonic growth in the seven epochs, but in the higher time resolution of the simulations, there are four `starbursts', in particular, a strong peak in the model at $t\approx0.8$~Gyr triggered by a high mass-ratio, gas-rich merger.  While this does cause a substantial jump in the integrated stellar mass in the $z=6$ interval, this galaxy did exceed the $10^8\mathrm{M}_\odot$ detection threshold in the $z=8$ bin, causing it to be classified as `non-burst'. The remaining panels of Figure~\ref{fig:smooths} again show more examples of both agreement and disagreement in classification. We find that only 46\% of galaxies classified into the `non-burst' category are free from bursts (following our definition based on the theoretical star formation histories) in the $z=7$ or $z=6$ bins.

It is informative to consider the ratio of stellar mass formed in the $z=7$ or $z=6$ bins to that formed in the earlier bins. For our `burst' sample, this quantity has a mean of 4.4 with a standard deviation of 11.8, while the `non-burst' sample has a mean of 2.4 and a standard deviation of 1.9. This difference in means is expected---by construction, the galaxies in the `burst' sample are those which have significantly more star formation in the latest two redshift bins than in earlier ones. However, the fact that the mean for the `non-burst' sample is still greater than 1 indicates that, even in these `non-burst' galaxies, the star formation rate is typically rising at these times, as can be clearly seen in Figure~\ref{fig:smooths}. The difference in standard deviations is also noteworthy. The `burst' galaxies show a much larger standard deviation, consistent with them more often (but not always) being dominated by a single peak of star formation in the latest two redshift bins.

Figures~\ref{fig:bursts} and \ref{fig:smooths} show that not all mergers trigger starbursts, which require both enough gas to fuel a substantial amount of star formation and to form those stars rapidly, which simulations show means it must be added to the dense spheroid component of the galaxy. In our model, major mergers always disrupt the existing galaxy, leaving a dense spheroid, but they do not always have sufficient gas to result in a starburst.  In Figure~\ref{fig:bursts}, it can be seen that major mergers (indicated by star symbols) almost always trigger a clear burst of star formation. Minor mergers do not disrupt pre-existing disks, but can deposit gas from a satellite into the dense spheroid of the central, resulting in enhanced star formation. In Figure~\ref{fig:bursts}, there are several instances where a minor merger (indicated by circle symbols) does result in a spike in star formation rate, resulting from the deposition of gas into the spheroid of the central galaxy. We find that starbursts, as identified in Figures~\ref{fig:bursts} and \ref{fig:smooths}) contribute 22\% of 26\% of the total stellar mass formed in our ``burst'' and `non-burst' samples respectively, which means that starbursts are not the proximate cause of whether a star formation history is classified as a \emph{burst} or \emph{non-burst} in the \citetalias{2024ApJ...964..150D} categories.

Finally, we note that some galaxies are very close to the boundary of our `burst'/`non-burst' classification. For example, the galaxy shown in the middle left panel of Figure~\ref{fig:bursts} forms a stellar mass in the $z\approx 8$ bin that is just below the $10^8\mathrm{M}_\odot$ threshold (i.e., the green dot at $t\approx 0.6$~Gyr lies just below the dotted line). If just slightly more stellar mass had formed in this redshift bin, this galaxy would have been classified as a non-burst. The opposite case (where a galaxy would have been classified as a burst had it formed just slightly less stellar mass in this redshift bin) can be seen in the top center panel of Figure~\ref{fig:smooths}. In reality, observations of such galaxies, which would inevitably have noisy measurements of the stellar mass in each redshift bin, would sometimes classify such galaxies as `bursts', sometimes as non-bursts, making them ambiguous.

Based on this discussion of `burst' and `starbursts', we conclude that the half of the galaxies classified as `bursts' are not the product of a few large \emph{starbursts}.  This suggests that the rapid growth in stellar mass seen for $12<z<6$ is the result of intense, galaxy-wide star formation, such as might be expected in the early gas-rich environment. We may call them `bursts' because of their rapidity, but most of that growth must be due to gas infall or acquisition that takes place at a rate that exceeds anything seen in later epochs.  What dominates the growth of the first galaxies might be called \emph{galaxy-bursts}. Based on mass growth in the `7' to `6' redshifts, we think that---
if the star formation rate increases by a factor of 2--3, and for a period of time that is a factor 3 longer than that in the redshift 7 and 6 bins combined---the by $z=3$ stellar masses could have increased by a factor of 6--9. The $z\sim3$ galaxies descending from Figure~\ref{SFH-1Gyr} would grow to between $10^{9}$\Msun and $10^{10.4}$\Msun, which seems consistent with observations.

\begin{figure*}
    \centering
    \begin{tabular}{cc}
    \includegraphics[width=0.5 \linewidth]{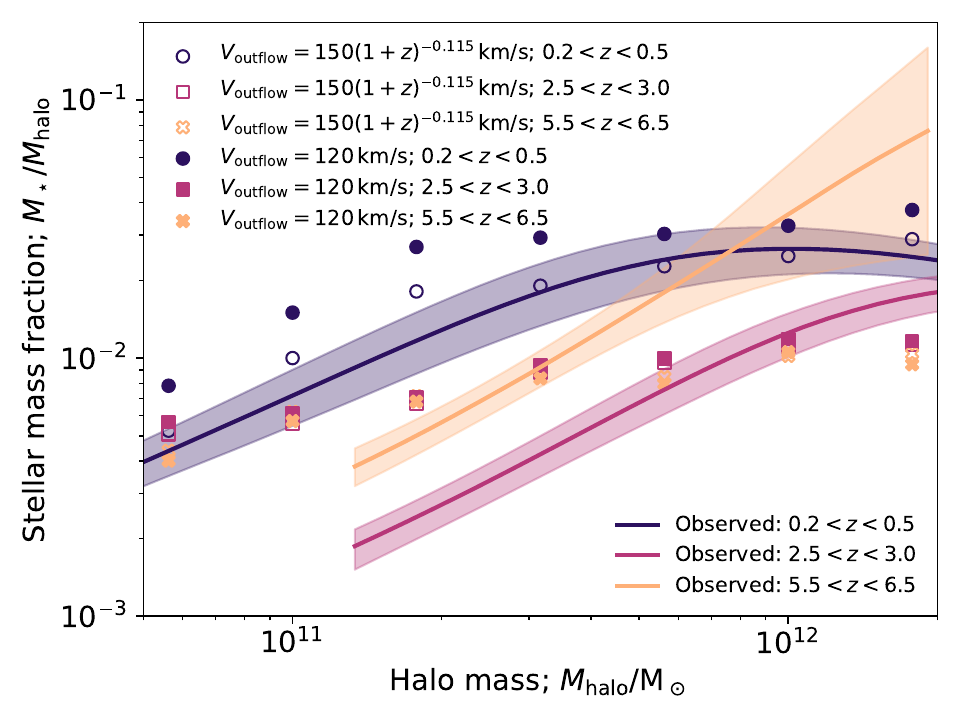} &  \includegraphics[width=0.5 \linewidth]{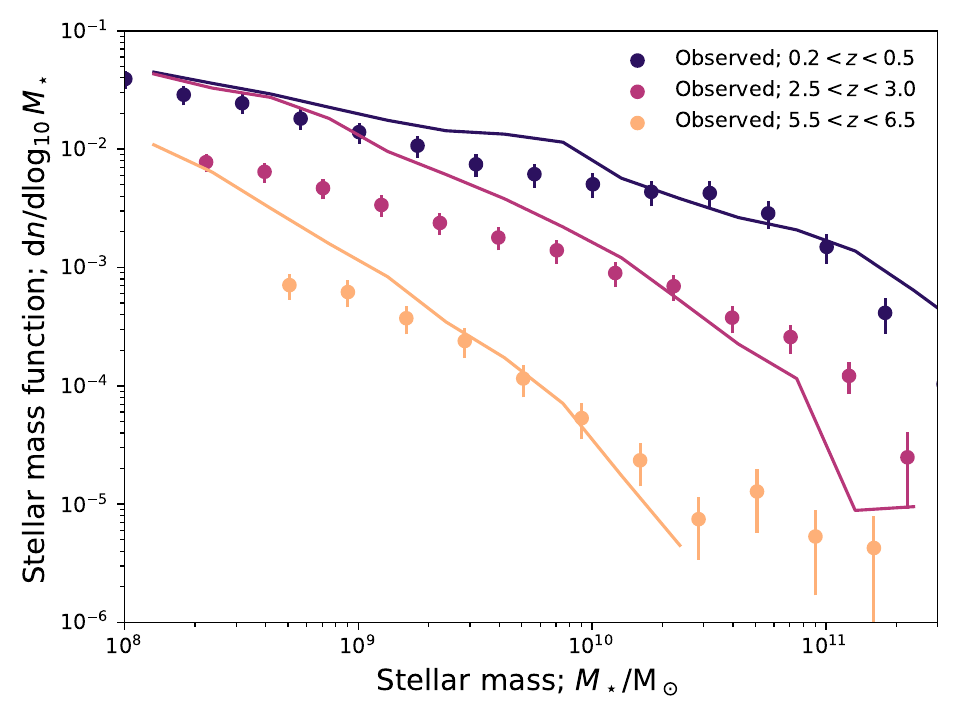}
    \end{tabular}
    \caption{\emph{Left panel:} The stellar mass--halo mass relation at three different redshift ranges: $z=0.2$--0.5 (blue), $z=2.5$--3.0 (purple), and $z=5.5$--6.0 (orange). Solid lines indicate the median relation inferred from observed galaxies by \cite{2025A&A...695A..20S}, with the shaded regions indicating the $1\sigma$ confidence interval on this median. Filled points show the results from our {\sc Galacticus} model that matches the assembly of stellar mass at $z > 6$. Open circles show a variant of this model in which we set $V_\mathrm{outflow} = 150 (1+z)^{-0.115}$~km/s. \emph{Right panel: The stellar mass function at the same three redshift ranges. Points with error bars show observational results from \cite{2025A&A...695A..20S}, while lines show results from our variant {\sc Galacticus} model (i.e., that shown by open symbols in the left panel).}}
    \label{fig:SHMR}
\end{figure*}

\section{Later Growth}

Our original model was calibrated to approximately match the properties of galaxies at $z=0$. Having reduced the strength of feedback to allow us to match the properties of galaxies at high redshifts, we may expect that our model will now produce too much star formation by $z=0$. To explore this, we run our model to $z=0$, outputting results at $z=2.75$ and $z=0.35$ to compare the recent observational determination of the stellar mass--halo mass relation from \cite{2025A&A...695A..20S}. The left panel of Figure~\ref{fig:SHMR} shows the stellar mass to halo mass ratio as a function of halo mass for three redshift ranges: $z=0.2$--0.5 (blue), $z=2.5$--3.0 (purple), and $z=5.5$--6.0 (orange). Solid lines indicate the median relation inferred from observed galaxies by \cite{2025A&A...695A..20S}, with the shaded regions indicating the $1\sigma$ confidence interval on this median. Points show the results from our {\sc Galacticus} model with feedback tuned to match the results shown in Figure~\ref{fig:stellarMass}.

At $z\approx 6$, galaxies with stellar masses of $10^8\mathrm{M}_\odot$ and greater, as considered in this work, are typically found in halos of masses $M_\mathrm{halo} \gtrsim 10^{11}\mathrm{M}_\odot$). At $M_\mathrm{halo} = 10^{11}\mathrm{M}_\odot$, our model moderately over-predicts the median stellar mass to halo mass ratio, by around 50\%, but agrees closely with the inferred ratio for $M_\mathrm{halo} \approx 2$--$3\times10^{11}\mathrm{M}_\odot$. At higher halo masses, at $z\approx 6$, our model substantially under-predicts the inferred ratio---although at these masses the observational uncertainties become quite large. This result is, therefore, consistent with that shown in Figure~\ref{fig:stellarMass} at $z=6$.

At $z=3$ our model significantly overpredicts the inferred stellar mass to halo mass ratio below $M_\mathrm{halo}=10^{12}\mathrm{M}_\odot$---by a factor of approximately 4 at $M_\mathrm{halo}=10^{11}\mathrm{M}_\odot$. By $z=0$, our model overpredicts the inferred ratio at almost all masses by factors of up to 2. The weaker feedback required in our model to match observations at high redshifts results in insufficient feedback at low redshifts, and an overproduction of stellar mass in the late-time universe.

To examine how much feedback would need to change to avoid this overproduction of stellar mass by $z=0$ we modify our feedback model by allowing the feedback parameter $V_\mathrm{outflow}$ to vary with redshift as
\begin{equation}
V_\mathrm{outflow}(z) = V_\mathrm{outflow} (1+z)^\gamma,
\end{equation}
where $\gamma$ is a new parameter. We find that choosing $V_\mathrm{outflow}=150$~km/s and $\gamma=-0.115$ (such that $V_\mathrm{outflow}(z=6) \approx 125$~km/s as before) leaves our predictions at $z=6$ approximately unchanged, while reducing the stellar masses of galaxies at $z=0$ to be much closer to expectations from the stellar mass--halo mass relation, as shown by the open circles in Figure~\ref{fig:SHMR}.

Lastly, we show in the right panel of Figure~\ref{fig:SHMR} the stellar mass function in the same three redshift intervals, comparing observations from \cite{2025A&A...695A..20S}, shown by points with error bars, to predictions from the variable feedback model described above. This demonstrates that, at $z=6$, not only has our model produced the correct total mass of stars, but has apportioned them into approximately the correct distribution of galaxy masses (although it does somewhat overpredict the $z=6$ stellar mass function at the lowest masses observed). Our model prediction for the stellar mass function in the $0.2 < z < 0.5$ interval is also in reasonable agreement with the data, although it overpredicts the stellar mass function at the highest masses (where AGN feedback, which we have not explored or attempted to calibrate in this work, may be relevant). Interestingly, at the intermediate redshift of $z=3$ our predicted stellar mass function is steeper than the observations.

\section{Discussion}
 
\begin{figure*}
\begin{center}
%\begin{tabular}{cc}
%\includegraphics[width=8cm]{galaxyMergerTree24.pdf}
%\includegraphics[width=8cm]{galaxyMergerTree2.pdf}
\includegraphics[width=\linewidth]{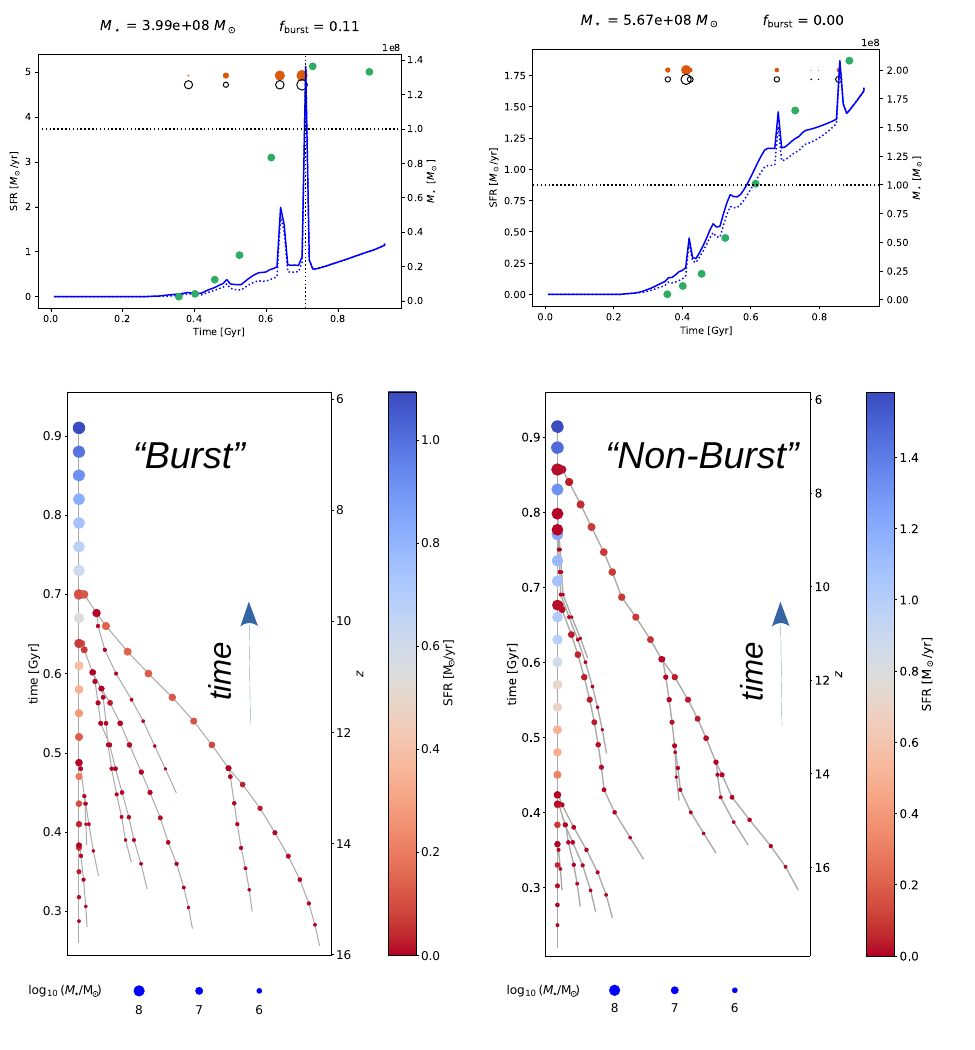}
%\end{tabular}
\end{center}
\caption{Comparison of stellar-mass-growth profiles (see Figures~\ref{fig:bursts} \& \ref{fig:smooths}) using \textsc{Galacticus} merger trees, for both a `burst' (left) and a `non-burst' galaxy (right). The merger trees show how each of these galaxies at $z\approx6$ was assembled through the mergers of many progenitors. In this figure, time increases upwards, and position on the $x$-axis is arbitrary. Each circle indicates a snapshot of a progenitor galaxy. Circle size is proportional to the stellar mass of the galaxy to the one-third power, while circle color indicates the star formation rate. Black lines connect progenitors across time and, where they converge, indicate merger events. \emph{Left panel:} In the `burst' example, progenitor galaxies are gradually forming stars up until just prior to $z\approx9$, at which point a merger at $t=0.7$~Gyr triggers a short-lived `starburst' of star formation fueled by gas delivered by the merging galaxy progenitors. After this `starburst' has completed, the star formation rate returns to its prior level, and slowly increases to $t=0.9$~Gyr, in a manner similar to that seen in the galaxy shown in the right panel. \emph{Right panel:} In the 'non-burst' example, the single galaxy observed at $z\sim10$ has been made by a steady merging and slow growth of only three progenitor galaxies, after which a high SFR is sustained through $z\sim6$ through the gas available from a growing halo mass.}
\label{fig:galaxyMergerTree}
\end{figure*}

We have compared the predictions from the {\sc Galacticus} model for the assembly of stellar mass in galaxies in the redshift range $z=6$--12 to observational estimates from \citetalias{2024ApJ...964..150D}. We emphasize that our model results are true \emph{predictions}, generated using a model previously calibrated to lower redshift data. Making testable predictions is, of course, the only way to quantify the scientific utility of a model. Our prior model accurately predicts the \emph{shape} of stellar mass growth in galaxies with stellar masses $M_\star \gtrsim 10^8\mathrm{M}_\odot$ across this whole redshift range. However, it under-predicts the \emph{total stellar mass} in these galaxies by a factor of 3--4. We find that reducing the strength of feedback in our model allows it to produce an excellent match to the observed stellar mass as a function of redshift. We also find that a model with very weak feedback dramatically overproduces stellar mass, and that models in which the stellar mass to halo mass ratio is a fixed fraction above some minimum halo mass can not reproduce the trend with redshift that is observed. Together, these results highlight the crucial importance of feedback at high redshifts.

We classified model galaxies into `burst' and `non-burst' classes following the observationally-motivated method of \citetalias{2024ApJ...964..150D}, finding generally good agreement with the fraction of galaxies in each class when compared with the observational data. 
Using the simulations to investigate with higher time resolution, we find that \emph{both} types make only $25\%$ of their stars in 30--40~Myr `starbursts' as are common at later epochs, so the main mode of star of stellar mass growth is likely galaxy scale and representative of all galaxies at the $12<z<6$ epoch of first growth.

Figure~\ref{fig:galaxyMergerTree} illustrates how the presence or absence of `starbursts' can alter the classification of the \citetalias{2024ApJ...964..150D} burst classes. To help see this, we compare the growth histories for two of our model galaxies---a `burst' and a `non-burst' galaxy. On top is the star-formation growth history of each galaxy in the manner of Figures~\ref{fig:bursts} \& \ref{fig:smooths}, and below each is their merger tree extracted from \textsc{Galacticus}. The merger trees resolve their assemblies from smaller systems at $z>12$ and $M<10^8$\Msun---showing their masses, star-formation rates, and gas content.

While merger trees are typically made for dark matter halos, Figure~\ref{fig:galaxyMergerTree} illustrates the merging of the progenitor galaxies (baryons) that merge to form a single galaxy (the track along the y-axis) of stellar mass $M_\star \sim 10^9$\Msun\ at $z\approx6$ for both examples. The circles are a snapshot of progenitor galaxies at each redshift with time (running up the figure---note that position on the $x$-axis is arbitrary) that lead to mergers between progenitor systems. Circles are colored by the star formation rate (as shown by the color bar in the figure), and their size is scaled in proportion to their stellar mass, as shown at the bottom.

As seen in the top left `history' panel, this is a galaxy with a gas-rich merger at 0.7~Gyr. The merger tree shows that the star formation rates remained low (see the line along the y-axis) in the many small progenitors, until a merger at $t=0.7$~Gyr (seen in both the upper and lower panels) triggers a short-lived `starburst' of star formation fueled by gas delivered by the merging galaxy progenitors. After this `starburst' has completed, the star formation rate returns to its prior level, and slowly increases to $t=0.9$~Gyr, in a manner similar to that seen in the galaxy shown in the right panel. In this case, it is the presence of this `starburst' which causes this galaxy to be placed into the `burst' class. 

In contrast, the `history' in the top right panel shows a galaxy that grows steadily with a smoothly increasing star formation rate, with only small enhancements in star formation rate due to mergers. The merger tree below shows only many gas-poor minor mergers at later epochs. The star formation rate increases steadily in the main branch, in agreement with the top `history' of a steadily growing star formation rate and stellar mass, and no `starbursts' from a sudden influx of gas. Given the complexity of these two examples, it is clear that the division of `burst' and `non-burst' is not `clean,' and that it is strongly modulated by the amount of gas delivered by the progenitors as assembly progresses.  

It is important to note that this complexity is present in the stellar mass assembly history---the additional complexities that arise in the conversion of this into an observable spectrum of the galaxy can only exacerbate the confusion. As such, it is important that models first verify that they achieve the correct mass growth in galaxies before comparing directly with broad-band luminosities or spectra. Two spectra that appear very different may be distinguished by only a very short timescale `starburst` and their overall mass growth may, in fact, be quite similar.

Finally, we note that the reduced strength in feedback required by our model to match observations at $z>6$ results in an overproduction of stellar mass at lower redshifts---by around a factor of 2 across a wide range of halo masses at $z=0$. This suggests that constructing a model that will produce the correct relation of stellar mass-to-halo mass across all redshifts will require a more complex feedback model. Our current model is quite simplistic: it assumes a rate of energy input proportional to the instantaneous star formation rate, and a mass loading factor that scales with the characteristic circular velocity of each galaxy (a measure of the depth of the gravitational potential well). A simple change to this model---allowing a weak dependence of the characteristic velocity on redshift ($V_\mathrm{outflow}(z) = 150 (1+z)^{-0.115}$ km s$^{-1}$ such that $V_\mathrm{outflow}(z=6)\approx 125$~km s$^{-1}$ (matching the reduced strength feedback model used in this work) allows for a better match to $z=0$ stellar masses, as shown by the open symbols in Figure~\ref{fig:SHMR}. More realistic models of feedback would take into account detailed energy input rates from stellar populations of different ages (\eg energy input from type Ia supernovae) will be delayed relative to star formation by timescales that are significant at $z\gtrsim6$, would help reduce feedback at those redshifts and incorporate dependencies on other galaxy properties such as gas surface density or gas fraction \citep{creasey_how_2013}.

The title of \citetalias{2024ApJ...964..150D} suggested that star formation histories of the first galaxies were `burst' dominated as deduced in data with poor time resolution.  The \textsc{Galacticus} simulations allow us to revisit that claim because, despite this difference, the model and data agree in fundamental properties.  With this perspective, we reconsider whether `burst' is a proper description for roughly half the population that is so identified.  It is fair to say that, from the perspective what has traditionally been described as a `starburst'---sudden episodes of star formation on time scales of $10^{7}$ yr, the histories are not dominated by these, but by very high rates of star formation---on a \emph{galactic} scale---fueled by mergers or by the accretion of large, mostly-gas clouds that might be much more common at these epochs.

To reiterate our conclusions from this comparison of theory (\textsc{Galacticus}) and observations (\emph{JWST}) for stellar mass growth in the first galaxies:
\begin{enumerate}
    \item the shape of the stellar mass growth curve with redshift is matched by \textsc{Galacticus} without `tuning,' and its normalization can be matched by adjusting the strength of feedback in the model---inviting the question of if this matches expectations of detailed theoretical models of feedback;
    \item using the same `burst' criteria as \citetalias{2024ApJ...964..150D}, the fraction ($\sim 50\%$) of galaxies with an apparent burst in star formation matches observations;
    \item based on these successes, we infer that starbursts ($\sim$30~Myr timescale) seen in the simulations must also occur in these early-epoch galaxies, but these provide only about one quarter of total star formation, which suggests high and increasing rates of star formation over the entire galaxy on timescales of hundreds of Myr;
    \item a continuing, modest increase in star formation rates until $z=3$ should be able to match observational estimates of stellar masses of galaxies at that epoch;
    \item matching the evolution of the stellar mass--halo mass relation and stellar mass function requires a feedback efficiency that evolves slowly with redshift.
\end{enumerate}

This paper demonstrates that establishing the mass growth in the first galaxies provides a much better diagnosis of their development than spectral properties alone. Much larger data sets of the kind used here, combined with higher spectral resolution data, will be required to constrain these histories and the stellar population differences over all of cosmic history. Fortunately, present facilities are capable of gathering these data with the needed variety and accuracy, so expectations are high for a much better understanding of how the first galaxies grew into the galaxies of today.

\begin{acknowledgments}
AD and AB acknowledge the Carnegie Institution for Science for their generous and faithful support of our science programs. Calculations in this work were carried out on the OBS HPC compute resources provided by the Carnegie Institution for Science.  The authors thank Mike Boylan-Kolchin and Robert Feldman for providing results from the \emph{FIREbox} simulations. 
\end{acknowledgments}

\begin{contribution}
%%This section gives authors the space to recognize author contributions. The text inside this environment is NOT counted towards the total word quanta. At a minimum, manuscripts are expected to include this text:

All authors contributed equally to this work.

%% But authors are expected to provide more specific details, e.g. 
%%
%%SC was responsible for writing and submitting the manuscript.
%%WWM came up with the initial research concept and edited the manuscript.
%%OTS obtained the funding and edited the manuscript.
%%EBF provided the formal analysis and validation. He also edited the manuscript.
%%GEH Supervised the undergraduates, wrote the software and administers the project github and Zenodo repositories.
%%
%% Authors can use the Contributor Role Taxonomy (CRediT) at
%% https://credit.niso.org
%% for ideas on how write a good statement tailored to their needs.

\end{contribution}

%% To help institutions obtain information on the effectiveness of their 
%% telescopes the AAS Journals has created a group of keywords for telescope 
%% facilities.
%
%% Following the acknowledgments section, use the following syntax and the
%% \facility{} or \facilities{} macros to list the keywords of facilities used 
%% in the research for the paper.  Each keyword is check against the master 
%% list during copy editing.  Individual instruments can be provided in 
%% parentheses, after the keyword, but they are not verified.
\facilities{HST(STIS), JWST(STIS), Swift(XRT and UVOT), AAVSO, CTIO:1.3m, CTIO:1.5m, CXO}

%% Similar to \facility{}, there is the optional \software command to allow 
%% authors a place to specify which programs were used during the creation of 
%% the manuscript. Authors should list each code and include either a
%% citation or url to the code inside ()s when available.
\software{astropy \citep{2013A&A...558A..33A,2018AJ....156..123A,2022ApJ...935..167A},  
          Cloudy \citep{2013RMxAA..49..137F}, 
          Source Extractor \citep{1996A&AS..117..393B}
          }
          
%% Appendix material should be preceded with a single \appendix command.
%% There should be a \section command for each appendix. Mark appendix
%% subsections with the same markup you use in the main body of the paper.
%%
%% Each Appendix (indicated with \section) will be lettered A, B, C, etc.
%% The equation counter will reset when it encounters the \appendix
%% command and will number appendix equations (A1), (A2), etc. The
%% Figure and Table counter will not reset.

\bibliographystyle{aasjournal}
\bibliography{sample7}{}

@ARTICLE{2024ApJ...964..150D,
       author = {{Dressler}, Alan and {Rieke}, Marcia and {Eisenstein}, Daniel and {Stark}, Daniel P. and {Burns}, Chris and {Bhatawdekar}, Rachana and {Bonaventura}, Nina and {Boyett}, Kristan and {Bunker}, Andrew J. and {Carniani}, Stefano and {Charlot}, Stephane and {Hausen}, Ryan and {Misselt}, Karl and {Tacchella}, Sandro and {Willmer}, Christopher},
        title = "{Building the First Galaxies{\textemdash}Chapter 2. Starbursts Dominate the Star Formation Histories of 6 < z < 12 Galaxies}",
      journal = {\apj},
         year = 2024,
        month = apr,
       volume = {964},
       number = {2},
          eid = {150},
        pages = {150},
          doi = {10.3847/1538-4357/ad1923},
archivePrefix = {arXiv},
       eprint = {2306.02469},
 primaryClass = {astro-ph.GA},
       adsurl = {https://ui.adsabs.harvard.edu/abs/2024ApJ...964..150D}
}

@ARTICLE{2023ApJ...947...L27,
       author = {{Dressler}, Alan and {Vulcani}, Benedetta and {Treu}, Tommaso and {Rieke}, Marcia and {Burns}, Chris and {Calabr{\`o}}, Antonello and {Bonchi}, Andrea and {Castellano}, Marco and {Fontana}, Adriano and {Leethochawalit}, Nicha and {Mason}, Charlotte and {Merlin}, Emiliano and {Morishita}, Takahiro and {Paris}, Diego and {Bradac}, Marusa and {Mercurio}, Amata and {Nanayakkara}, Themiya and {Poggianti}, Bianca M. and {Santini}, Paola and {Wang}, Xin and {Misselt}, Karl and {Stark}, Daniel P. and {Willmer}, Christopher},
        title = "{Early Results from GLASS-JWST. XVII. Building the First Galaxies-Chapter 1. Star Formation Histories for 5 < z < 7 Galaxies}",
      journal = {\apjl},
         year = 2023,
        month = apr,
       volume = {947},
       number = {2},
          eid = {L27},
        pages = {L27},
          doi = {10.3847/2041-8213/ac9ebb},
archivePrefix = {arXiv},
       eprint = {2208.04292},
 primaryClass = {astro-ph.GA},
       adsurl = {https://ui.adsabs.harvard.edu/abs/2023ApJ...947L..27D}
}

@MISC{1996STIN...9713870D,
       author = {{Dressler}, Alan and {Brown}, Robert A. and {Davidsen}, Arthur F. and {Ellis}, Richard S. and {Freedman}, Wendy L. and {Green}, Richard F. and {Hauser}, Michael G. and {Kirshner}, Robert P. and {Kulkarni}, Shrinivas and {Lilly}, Simon J. and {Margon}, Bruce H. and {Porco}, Carolyn C. and {Richstone}, Douglas O. and {Stockman}, H.~S. and {Thronson}, Jr., Harley A. and {Tonry}, John L. and {Truran}, James and {Weiler}, Edward J.},
        title = "{Exploration an the Search for Origins: A Vision for Ultraviolet-Optical-Infrared Space Astronomy}",
 howpublished = {Technical Report, HST \& Beyond Committee},
         year = 1996,
        month = may,
        pages = {13870},
       adsurl = {https://ui.adsabs.harvard.edu/abs/1996STIN...9713870D}
}

@ARTICLE{2022ApJ...935..167A,
       author = {{Astropy Collaboration} and {Price-Whelan}, Adrian M. and {Lim}, Pey Lian and {Earl}, Nicholas and {Starkman}, Nathaniel and {Bradley}, Larry and {Shupe}, David L. and {Patil}, Aarya A. and {Corrales}, Lia and {Brasseur}, C.~E. and {N{\"o}the}, Maximilian and {Donath}, Axel and {Tollerud}, Erik and {Morris}, Brett M. and {Ginsburg}, Adam and {Vaher}, Eero and {Weaver}, Benjamin A. and {Tocknell}, James and {Jamieson}, William and {van Kerkwijk}, Marten H. and {Robitaille}, Thomas P. and {Merry}, Bruce and {Bachetti}, Matteo and {G{\"u}nther}, H. Moritz and {Aldcroft}, Thomas L. and {Alvarado-Montes}, Jaime A. and {Archibald}, Anne M. and {B{\'o}di}, Attila and {Bapat}, Shreyas and {Barentsen}, Geert and {Baz{\'a}n}, Juanjo and {Biswas}, Manish and {Boquien}, M{\'e}d{\'e}ric and {Burke}, D.~J. and {Cara}, Daria and {Cara}, Mihai and {Conroy}, Kyle E. and {Conseil}, Simon and {Craig}, Matthew W. and {Cross}, Robert M. and {Cruz}, Kelle L. and {D'Eugenio}, Francesco and {Dencheva}, Nadia and {Devillepoix}, Hadrien A.~R. and {Dietrich}, J{\"o}rg P. and {Eigenbrot}, Arthur Davis and {Erben}, Thomas and {Ferreira}, Leonardo and {Foreman-Mackey}, Daniel and {Fox}, Ryan and {Freij}, Nabil and {Garg}, Suyog and {Geda}, Robel and {Glattly}, Lauren and {Gondhalekar}, Yash and {Gordon}, Karl D. and {Grant}, David and {Greenfield}, Perry and {Groener}, Austen M. and {Guest}, Steve and {Gurovich}, Sebastian and {Handberg}, Rasmus and {Hart}, Akeem and {Hatfield-Dodds}, Zac and {Homeier}, Derek and {Hosseinzadeh}, Griffin and {Jenness}, Tim and {Jones}, Craig K. and {Joseph}, Prajwel and {Kalmbach}, J. Bryce and {Karamehmetoglu}, Emir and {Ka{\l}uszy{\'n}ski}, Miko{\l}aj and {Kelley}, Michael S.~P. and {Kern}, Nicholas and {Kerzendorf}, Wolfgang E. and {Koch}, Eric W. and {Kulumani}, Shankar and {Lee}, Antony and {Ly}, Chun and {Ma}, Zhiyuan and {MacBride}, Conor and {Maljaars}, Jakob M. and {Muna}, Demitri and {Murphy}, N.~A. and {Norman}, Henrik and {O'Steen}, Richard and {Oman}, Kyle A. and {Pacifici}, Camilla and {Pascual}, Sergio and {Pascual-Granado}, J. and {Patil}, Rohit R. and {Perren}, Gabriel I. and {Pickering}, Timothy E. and {Rastogi}, Tanuj and {Roulston}, Benjamin R. and {Ryan}, Daniel F. and {Rykoff}, Eli S. and {Sabater}, Jose and {Sakurikar}, Parikshit and {Salgado}, Jes{\'u}s and {Sanghi}, Aniket and {Saunders}, Nicholas and {Savchenko}, Volodymyr and {Schwardt}, Ludwig and {Seifert-Eckert}, Michael and {Shih}, Albert Y. and {Jain}, Anany Shrey and {Shukla}, Gyanendra and {Sick}, Jonathan and {Simpson}, Chris and {Singanamalla}, Sudheesh and {Singer}, Leo P. and {Singhal}, Jaladh and {Sinha}, Manodeep and {Sip{\H{o}}cz}, Brigitta M. and {Spitler}, Lee R. and {Stansby}, David and {Streicher}, Ole and {{\v{S}}umak}, Jani and {Swinbank}, John D. and {Taranu}, Dan S. and {Tewary}, Nikita and {Tremblay}, Grant R. and {de Val-Borro}, Miguel and {Van Kooten}, Samuel J. and {Vasovi{\'c}}, Zlatan and {Verma}, Shresth and {de Miranda Cardoso}, Jos{\'e} Vin{\'\i}cius and {Williams}, Peter K.~G. and {Wilson}, Tom J. and {Winkel}, Benjamin and {Wood-Vasey}, W.~M. and {Xue}, Rui and {Yoachim}, Peter and {Zhang}, Chen and {Zonca}, Andrea and {Astropy Project Contributors}},
        title = "{The Astropy Project: Sustaining and Growing a Community-oriented Open-source Project and the Latest Major Release (v5.0) of the Core Package}",
      journal = {\apj},
         year = 2022,
        month = aug,
       volume = {935},
       number = {2},
          eid = {167},
        pages = {167},
          doi = {10.3847/1538-4357/ac7c74},
archivePrefix = {arXiv},
       eprint = {2206.14220},
 primaryClass = {astro-ph.IM},
       adsurl = {https://ui.adsabs.harvard.edu/abs/2022ApJ...935..167A}
}

@ARTICLE{2018AJ....156..123A,
       author = {{Astropy Collaboration} and {Price-Whelan}, A.~M. and {Sip{\H{o}}cz}, B.~M. and {G{\"u}nther}, H.~M. and {Lim}, P.~L. and {Crawford}, S.~M. and {Conseil}, S. and {Shupe}, D.~L. and {Craig}, M.~W. and {Dencheva}, N. and {Ginsburg}, A. and {VanderPlas}, J.~T. and {Bradley}, L.~D. and {P{\'e}rez-Su{\'a}rez}, D. and {de Val-Borro}, M. and {Aldcroft}, T.~L. and {Cruz}, K.~L. and {Robitaille}, T.~P. and {Tollerud}, E.~J. and {Ardelean}, C. and {Babej}, T. and {Bach}, Y.~P. and {Bachetti}, M. and {Bakanov}, A.~V. and {Bamford}, S.~P. and {Barentsen}, G. and {Barmby}, P. and {Baumbach}, A. and {Berry}, K.~L. and {Biscani}, F. and {Boquien}, M. and {Bostroem}, K.~A. and {Bouma}, L.~G. and {Brammer}, G.~B. and {Bray}, E.~M. and {Breytenbach}, H. and {Buddelmeijer}, H. and {Burke}, D.~J. and {Calderone}, G. and {Cano Rodr{\'\i}guez}, J.~L. and {Cara}, M. and {Cardoso}, J.~V.~M. and {Cheedella}, S. and {Copin}, Y. and {Corrales}, L. and {Crichton}, D. and {D'Avella}, D. and {Deil}, C. and {Depagne}, {\'E}. and {Dietrich}, J.~P. and {Donath}, A. and {Droettboom}, M. and {Earl}, N. and {Erben}, T. and {Fabbro}, S. and {Ferreira}, L.~A. and {Finethy}, T. and {Fox}, R.~T. and {Garrison}, L.~H. and {Gibbons}, S.~L.~J. and {Goldstein}, D.~A. and {Gommers}, R. and {Greco}, J.~P. and {Greenfield}, P. and {Groener}, A.~M. and {Grollier}, F. and {Hagen}, A. and {Hirst}, P. and {Homeier}, D. and {Horton}, A.~J. and {Hosseinzadeh}, G. and {Hu}, L. and {Hunkeler}, J.~S. and {Ivezi{\'c}}, {\v{Z}}. and {Jain}, A. and {Jenness}, T. and {Kanarek}, G. and {Kendrew}, S. and {Kern}, N.~S. and {Kerzendorf}, W.~E. and {Khvalko}, A. and {King}, J. and {Kirkby}, D. and {Kulkarni}, A.~M. and {Kumar}, A. and {Lee}, A. and {Lenz}, D. and {Littlefair}, S.~P. and {Ma}, Z. and {Macleod}, D.~M. and {Mastropietro}, M. and {McCully}, C. and {Montagnac}, S. and {Morris}, B.~M. and {Mueller}, M. and {Mumford}, S.~J. and {Muna}, D. and {Murphy}, N.~A. and {Nelson}, S. and {Nguyen}, G.~H. and {Ninan}, J.~P. and {N{\"o}the}, M. and {Ogaz}, S. and {Oh}, S. and {Parejko}, J.~K. and {Parley}, N. and {Pascual}, S. and {Patil}, R. and {Patil}, A.~A. and {Plunkett}, A.~L. and {Prochaska}, J.~X. and {Rastogi}, T. and {Reddy Janga}, V. and {Sabater}, J. and {Sakurikar}, P. and {Seifert}, M. and {Sherbert}, L.~E. and {Sherwood-Taylor}, H. and {Shih}, A.~Y. and {Sick}, J. and {Silbiger}, M.~T. and {Singanamalla}, S. and {Singer}, L.~P. and {Sladen}, P.~H. and {Sooley}, K.~A. and {Sornarajah}, S. and {Streicher}, O. and {Teuben}, P. and {Thomas}, S.~W. and {Tremblay}, G.~R. and {Turner}, J.~E.~H. and {Terr{\'o}n}, V. and {van Kerkwijk}, M.~H. and {de la Vega}, A. and {Watkins}, L.~L. and {Weaver}, B.~A. and {Whitmore}, J.~B. and {Woillez}, J. and {Zabalza}, V. and {Astropy Contributors}},
        title = "{The Astropy Project: Building an Open-science Project and Status of the v2.0 Core Package}",
      journal = {\aj},
         year = 2018,
        month = sep,
       volume = {156},
       number = {3},
          eid = {123},
        pages = {123},
          doi = {10.3847/1538-3881/aabc4f},
archivePrefix = {arXiv},
       eprint = {1801.02634},
 primaryClass = {astro-ph.IM},
       adsurl = {https://ui.adsabs.harvard.edu/abs/2018AJ....156..123A}
}

@ARTICLE{2013A&A...558A..33A,
       author = {{Astropy Collaboration} and {Robitaille}, Thomas P. and
         {Tollerud}, Erik J. and {Greenfield}, Perry and {Droettboom}, Michael and
         {Bray}, Erik and {Aldcroft}, Tom and {Davis}, Matt and
         {Ginsburg}, Adam and {Price-Whelan}, Adrian M. and
         {Kerzendorf}, Wolfgang E. and {Conley}, Alexander and {Crighton}, Neil and
         {Barbary}, Kyle and {Muna}, Demitri and {Ferguson}, Henry and
         {Grollier}, Fr{\'e}d{\'e}ric and {Parikh}, Madhura M. and
         {Nair}, Prasanth H. and {Unther}, Hans M. and {Deil}, Christoph and
         {Woillez}, Julien and {Conseil}, Simon and {Kramer}, Roban and
         {Turner}, James E.~H. and {Singer}, Leo and {Fox}, Ryan and
         {Weaver}, Benjamin A. and {Zabalza}, Victor and {Edwards}, Zachary I. and
         {Azalee Bostroem}, K. and {Burke}, D.~J. and {Casey}, Andrew R. and
         {Crawford}, Steven M. and {Dencheva}, Nadia and {Ely}, Justin and
         {Jenness}, Tim and {Labrie}, Kathleen and {Lim}, Pey Lian and
         {Pierfederici}, Francesco and {Pontzen}, Andrew and {Ptak}, Andy and
         {Refsdal}, Brian and {Servillat}, Mathieu and {Streicher}, Ole},
        title = "{Astropy: A community Python package for astronomy}",
      journal = {\aap},
         year = "2013",
        month = "Oct",
       volume = {558},
          eid = {A33},
        pages = {A33},
          doi = {10.1051/0004-6361/201322068},
archivePrefix = {arXiv},
       eprint = {1307.6212},
 primaryClass = {astro-ph.IM},
       adsurl = {https://ui.adsabs.harvard.edu/abs/2013A&A...558A..33A}
}

@ARTICLE{1996A&AS..117..393B,
       author = {{Bertin}, E. and {Arnouts}, S.},
        title = "{SExtractor: Software for source extraction.}",
      journal = {\aaps},
         year = "1996",
        month = "Jun",
       volume = {117},
        pages = {393-404},
          doi = {10.1051/aas:1996164},
       adsurl = {https://ui.adsabs.harvard.edu/abs/1996A&AS..117..393B}
}

@ARTICLE{2013RMxAA..49..137F,
       author = {{Ferland}, G.~J. and {Porter}, R.~L. and {van Hoof}, P.~A.~M. and
         {Williams}, R.~J.~R. and {Abel}, N.~P. and {Lykins}, M.~L. and
         {Shaw}, G. and {Henney}, W.~J. and {Stancil}, P.~C.},
        title = "{The 2013 Release of Cloudy}",
      journal = {\rmxaa},
         year = "2013",
        month = "Apr",
       volume = {49},
        pages = {137-163},
archivePrefix = {arXiv},
       eprint = {1302.4485},
 primaryClass = {astro-ph.GA},
       adsurl = {https://ui.adsabs.harvard.edu/abs/2013RMxAA..49..137F}
}

@ARTICLE{2024arXiv241011680D,
       author = {{Driskell}, Trey and {Nadler}, Ethan O. and {Benson}, Andrew and {Gluscevic}, Vera},
        title = "{Population synthesis and astrophysical inference for high-$z$ JWST galaxies}",
      journal = {arXiv e-prints},
         year = 2024,
        month = oct,
          eid = {arXiv:2410.11680},
        pages = {arXiv:2410.11680},
          doi = {10.48550/arXiv.2410.11680},
archivePrefix = {arXiv},
       eprint = {2410.11680},
 primaryClass = {astro-ph.GA},
       adsurl = {https://ui.adsabs.harvard.edu/abs/2024arXiv241011680D}
}

@ARTICLE{2018MNRAS.474.5206K,
       author = {{Knebe}, Alexander and {Stoppacher}, Doris and {Prada}, Francisco and {Behrens}, Christoph and {Benson}, Andrew and {Cora}, Sofia A. and {Croton}, Darren J. and {Padilla}, Nelson D. and {Ruiz}, Andr{\'e}s N. and {Sinha}, Manodeep and {Stevens}, Adam R.~H. and {Vega-Mart{\'\i}nez}, Cristian A. and {Behroozi}, Peter and {Gonzalez-Perez}, Violeta and {Gottl{\"o}ber}, Stefan and {Klypin}, Anatoly A. and {Yepes}, Gustavo and {Enke}, Harry and {Libeskind}, Noam I. and {Riebe}, Kristin and {Steinmetz}, Matthias},
        title = "{MULTIDARK-GALAXIES: data release and first results}",
      journal = {\mnras},
         year = 2018,
        month = mar,
       volume = {474},
       number = {4},
        pages = {5206-5231},
          doi = {10.1093/mnras/stx2662},
archivePrefix = {arXiv},
       eprint = {1710.08150},
 primaryClass = {astro-ph.GA},
       adsurl = {https://ui.adsabs.harvard.edu/abs/2018MNRAS.474.5206K}
}

@ARTICLE{2012NewA...17..175B,
       author = {{Benson}, Andrew J.},
        title = "{G ALACTICUS: A semi-analytic model of galaxy formation}",
      journal = {\na},
         year = 2012,
        month = feb,
       volume = {17},
       number = {2},
        pages = {175-197},
          doi = {10.1016/j.newast.2011.07.004},
archivePrefix = {arXiv},
       eprint = {1008.1786},
 primaryClass = {astro-ph.CO},
       adsurl = {https://ui.adsabs.harvard.edu/abs/2012NewA...17..175B}
}

@ARTICLE{2015ARA&A..53...51S,
       author = {{Somerville}, Rachel S. and {Dav{\'e}}, Romeel},
        title = "{Physical Models of Galaxy Formation in a Cosmological Framework}",
      journal = {\araa},
         year = 2015,
        month = aug,
       volume = {53},
        pages = {51-113},
          doi = {10.1146/annurev-astro-082812-140951},
archivePrefix = {arXiv},
       eprint = {1412.2712},
 primaryClass = {astro-ph.GA},
       adsurl = {https://ui.adsabs.harvard.edu/abs/2015ARA&A..53...51S}
}

@ARTICLE{2025arXiv250117078S,
       author = {{Stark}, Daniel P. and {Topping}, Michael W. and {Endsley}, Ryan and {Tang}, Mengtao},
        title = "{Observations of the First Galaxies in the Era of JWST}",
      journal = {arXiv e-prints},
         year = 2025,
        month = jan,
          eid = {arXiv:2501.17078},
        pages = {arXiv:2501.17078},
          doi = {10.48550/arXiv.2501.17078},
archivePrefix = {arXiv},
       eprint = {2501.17078},
 primaryClass = {astro-ph.GA},
       adsurl = {https://ui.adsabs.harvard.edu/abs/2025arXiv250117078S}
}

@ARTICLE{2023NatAs...7..731B,
       author = {{Boylan-Kolchin}, Michael},
        title = "{Stress testing {\ensuremath{\Lambda}}CDM with high-redshift galaxy candidates}",
      journal = {Nature Astronomy},
         year = 2023,
        month = jun,
       volume = {7},
        pages = {731-735},
          doi = {10.1038/s41550-023-01937-7},
archivePrefix = {arXiv},
       eprint = {2208.01611},
 primaryClass = {astro-ph.CO},
       adsurl = {https://ui.adsabs.harvard.edu/abs/2023NatAs...7..731B}
}

@ARTICLE{2025MNRAS.536..988F,
       author = {{Feldmann}, Robert and {Boylan-Kolchin}, Michael and {Bullock}, James S. and {{\c{C}}atmabacak}, Onur and {Faucher-Gigu{\`e}re}, Claude-Andr{\'e} and {Hayward}, Christopher C. and {Kere{\v{s}}}, Du{\v{s}}an and {Lazar}, Alexandres and {Liang}, Lichen and {Moreno}, Jorge and {Oesch}, Pascal A. and {Quataert}, Eliot and {Shen}, Xuejian and {Sun}, Guochao},
        title = "{Elevated UV luminosity density at Cosmic Dawn explained by non-evolving, weakly mass-dependent star formation efficiency}",
      journal = {\mnras},
         year = 2025,
        month = jan,
       volume = {536},
       number = {1},
        pages = {988-1016},
          doi = {10.1093/mnras/stae2633},
archivePrefix = {arXiv},
       eprint = {2407.02674},
 primaryClass = {astro-ph.CO},
       adsurl = {https://ui.adsabs.harvard.edu/abs/2025MNRAS.536..988F}
}

@ARTICLE{2025arXiv250418618Y,
       author = {{Yung}, L.~Y. Aaron and {Somerville}, Rachel S. and {Iyer}, Kartheik G.},
        title = "{$Λ$CDM is still not broken: empirical constraints on the star formation efficiency at $z \sim 12-30$}",
      journal = {arXiv e-prints},
         year = 2025,
        month = apr,
          eid = {arXiv:2504.18618},
        pages = {arXiv:2504.18618},
          doi = {10.48550/arXiv.2504.18618},
archivePrefix = {arXiv},
       eprint = {2504.18618},
 primaryClass = {astro-ph.GA},
       adsurl = {https://ui.adsabs.harvard.edu/abs/2025arXiv250418618Y}
}

@ARTICLE{2025ApJ...982...23J,
       author = {{Jespersen}, Christian Kragh and {Steinhardt}, Charles L. and {Somerville}, Rachel S. and {Lovell}, Christopher C.},
        title = "{On the Significance of Rare Objects at High Redshift: The Impact of Cosmic Variance}",
      journal = {\apj},
         year = 2025,
        month = mar,
       volume = {982},
       number = {1},
          eid = {23},
        pages = {23},
          doi = {10.3847/1538-4357/adb422},
archivePrefix = {arXiv},
       eprint = {2403.00050},
 primaryClass = {astro-ph.GA},
       adsurl = {https://ui.adsabs.harvard.edu/abs/2025ApJ...982...23J}
}

@ARTICLE{2025ApJ...980...10J,
       author = {{Jeong}, Tae Bong and {Jeon}, Myoungwon and {Song}, Hyunmi and {Bromm}, Volker},
        title = "{Simulating High-redshift Galaxies: Enhancing UV Luminosity with Star Formation Efficiency and a Top-heavy IMF}",
      journal = {\apj},
         year = 2025,
        month = feb,
       volume = {980},
       number = {1},
          eid = {10},
        pages = {10},
          doi = {10.3847/1538-4357/ada27d},
archivePrefix = {arXiv},
       eprint = {2411.17007},
 primaryClass = {astro-ph.GA},
       adsurl = {https://ui.adsabs.harvard.edu/abs/2025ApJ...980...10J}
}

@ARTICLE{2025MNRAS.536.1018L,
       author = {{Lu}, Shengdong and {Frenk}, Carlos S. and {Bose}, Sownak and {Lacey}, Cedric G. and {Cole}, Shaun and {Baugh}, Carlton M. and {Helly}, John C.},
        title = "{A comparison of pre-existing {\ensuremath{\Lambda}}CDM predictions with the abundance of JWST galaxies at high redshift}",
      journal = {\mnras},
         year = 2025,
        month = jan,
       volume = {536},
       number = {1},
        pages = {1018-1034},
          doi = {10.1093/mnras/stae2646},
archivePrefix = {arXiv},
       eprint = {2406.02672},
 primaryClass = {astro-ph.GA},
       adsurl = {https://ui.adsabs.harvard.edu/abs/2025MNRAS.536.1018L}
}

@ARTICLE{2023MNRAS.522.3831F,
       author = {{Feldmann}, Robert and {Quataert}, Eliot and {Faucher-Gigu{\`e}re}, Claude-Andr{\'e} and {Hopkins}, Philip F. and {{\c{C}}atmabacak}, Onur and {Kere{\v{s}}}, Du{\v{s}}an and {Bassini}, Luigi and {Bernardini}, Mauro and {Bullock}, James S. and {Cenci}, Elia and {Gensior}, Jindra and {Liang}, Lichen and {Moreno}, Jorge and {Wetzel}, Andrew},
        title = "{FIREbox: simulating galaxies at high dynamic range in a cosmological volume}",
      journal = {\mnras},
         year = 2023,
        month = jul,
       volume = {522},
       number = {3},
        pages = {3831-3860},
          doi = {10.1093/mnras/stad1205},
archivePrefix = {arXiv},
       eprint = {2205.15325},
 primaryClass = {astro-ph.GA},
       adsurl = {https://ui.adsabs.harvard.edu/abs/2023MNRAS.522.3831F}
}

@ARTICLE{2025A&A...695A..20S,
       author = {{Shuntov}, M. and {Ilbert}, O. and {Toft}, S. and {Arango-Toro}, R.~C. and {Akins}, H.~B. and {Casey}, C.~M. and {Franco}, M. and {Harish}, S. and {Kartaltepe}, J.~S. and {Koekemoer}, A.~M. and {McCracken}, H.~J. and {Paquereau}, L. and {Laigle}, C. and {Bethermin}, M. and {Dubois}, Y. and {Drakos}, N.~E. and {Faisst}, A. and {Gozaliasl}, G. and {Gillman}, S. and {Hayward}, C.~C. and {Hirschmann}, M. and {Huertas-Company}, M. and {Jespersen}, C.~K. and {Jin}, S. and {Kokorev}, V. and {Lambrides}, E. and {Le Borgne}, D. and {Liu}, D. and {Magdis}, G. and {Massey}, R. and {McPartland}, C.~J.~R. and {Mercier}, W. and {McCleary}, J.~E. and {McKinney}, J. and {Oesch}, P.~A. and {Renzini}, A. and {Rhodes}, J.~D. and {Rich}, R.~M. and {Robertson}, B.~E. and {Sanders}, D. and {Trebitsch}, M. and {Tresse}, L. and {Valentino}, F. and {Vijayan}, A.~P. and {Weaver}, J.~R. and {Weibel}, A. and {Wilkins}, S.~M. and {Yang}, L.},
        title = "{COSMOS-Web: Stellar mass assembly in relation to dark matter halos across 0.2 < z < 12 of cosmic history}",
      journal = {\aap},
         year = 2025,
        month = mar,
       volume = {695},
          eid = {A20},
        pages = {A20},
          doi = {10.1051/0004-6361/202452570},
archivePrefix = {arXiv},
       eprint = {2410.08290},
 primaryClass = {astro-ph.GA},
       adsurl = {https://ui.adsabs.harvard.edu/abs/2025A&A...695A..20S}
}

@article{creasey_how_2013,
	title = {How supernova explosions power galactic winds},
	volume = {429},
	issn = {0035-8711},
	url = {http://adsabs.harvard.edu/abs/2013MNRAS.429.1922C},
	doi = {10.1093/mnras/sts439},
	urldate = {2014-03-03},
	journal = {Monthly Notices of the Royal Astronomical Society},
	author = {Creasey, Peter and Theuns, Tom and Bower, Richard G.},
	month = mar,
	year = {2013},
	pages = {1922--1948},
}

@ARTICLE{2014ApJ...796...75C,
       author = {{Chabrier}, Gilles and {Hennebelle}, Patrick and {Charlot}, St{\'e}phane},
        title = "{Variations of the Stellar Initial Mass Function in the Progenitors of Massive Early-type Galaxies and in Extreme Starburst Environments}",
      journal = {\apj},
         year = 2014,
        month = dec,
       volume = {796},
       number = {2},
          eid = {75},
        pages = {75},
          doi = {10.1088/0004-637X/796/2/75},
archivePrefix = {arXiv},
       eprint = {1409.8466},
 primaryClass = {astro-ph.GA},
       adsurl = {https://ui.adsabs.harvard.edu/abs/2014ApJ...796...75C}
}

@BOOK{2000eaa..book.....M,
       author = {{Murdin}, Paul},
        title = "{Encyclopedia of Astronomy and Astrophysics}",
         year = 2000,
       adsurl = {https://ui.adsabs.harvard.edu/abs/2000eaa..book.....M},
      publisher = {Institute of Physics Publishing, 2001}
}

%% This command is needed to show the entire author+affiliation list when
%% the collaboration and author truncation commands are used.  It has to
%% go at the end of the manuscript.
%\allauthors

%% Include this line if you are using the \added, \replaced, \deleted
%% commands to see a summary list of all changes at the end of the article.
%\listofchanges

\end{document}